\documentclass[a4paper,longnamesfirst,useAMS,usenatbib,usegraphicx,usedcolumn]{mn2e}

\usepackage{amsmath}
\usepackage{times}

\usepackage{url}
\usepackage{xr-hyper}
\usepackage{hyperref}

\shortcites{browne93,biggs02,tammann01,parodi00,press_etal,cohen00,biggs01,kochanek89,patnaik93,biggs99,castles02,koopmans03b}

\newcommand{\nt}{}
\newcommand{\papi}{Paper~I}
\newcommand{\papii}{Paper~II}

\newcommand{\mat}[1]{{\mathbfss{#1}}}

\newcommand{\zd}{z_{\rmn{d}}}
\newcommand{\dd}{d_{\rmn{d}}}
\newcommand{\ds}{d_{\rmn{s}}}
\newcommand{\dds}{d_{\rmn{ds}}}
\newcommand{\vc}[1]{\bmath{#1}}

\newcommand{\zs}{z_{\rmn{s}}}
\newcommand{\ma}{\mat{M}_{\rmn{A}}}
\newcommand{\mb}{\mat{M}_{\rmn{B}}}
\newcommand{\ra}{\vc r_{\rmn{A}}}
\newcommand{\rb}{\vc r_{\rmn{B}}}
\newcommand{\rs}{\bar{\vc r}_{\rmn{s}}}
\newcommand{\transp}[1]{{#1}^{\dagger}}
\newcommand{\re}{r_\epsilon}

\newcommand{\kmsmpc}{\ifmmode\text{km}\,\text{s}^{-1}\,\text{Mpc}^{-1}\else$\text{km}\,\text{s}^{-1}\,\text{Mpc}^{-1}$\fi}

\newcommand{\mymathstrut}{jkl'm}

\newcommand{\footstrut}{\vphantom{$g^{\mbox{A}}$}}

\newcommand{\sub}[1]{_{\mathrm{#1}}}

\newcommand{\mas}{{\ifmmode\text{mas}\else mas\fi}}

\newcommand{\BOii}{B0218+357}

\newcommand{\rtext}[1]{\quad\text{#1}}

\newcommand{\lc}{{\scshape LensClean}}

\newcommand{\lentil}{{\scshape LenTil}}

\newcommand{\clean}{{\scshape Clean}}

\newcommand{\difmap}{{\scshape DIFMAP}}

\newcommand{\hst}{{\it HST}}

\let\epsilon=\varepsilon

\newcommand{\pam}[2]{\begin{smallmatrix} #1 \\ #2\end{smallmatrix}}

\title[Models for the lens and source of B0218+357]{Models for the
  lens and source of B0218+357 \\ A \lc\ approach to determine $\bmath{H_0}$}
\author[O.~Wucknitz et al.]
       {O.~Wucknitz,$^{1,2,3}$\thanks{E-mail: {\tt
  olaf@astro.physik.uni-potsdam.de}} 
A.~D.~Biggs$^{1,4}$ and I.~W.~A.~Browne$^1$
  \\
$^1$ University of Manchester, Jodrell Bank Observatory, Macclesfield, Cheshire SK11 9DL, UK
         \\
$^2$ Hamburger Sternwarte, Universit{\"a}t Hamburg, Gojenbergsweg 112, 21029 Hamburg, Germany
\\
$^3$ Universit\"{a}t Potsdam,   Institut f\"{u}r Physik, Am Neuen Palais 10,
  14469 Potsdam, Germany \\
$^4$ Joint Institute for VLBI in Europe, Postbus 2, 7990 AA Dwingeloo, The
  Netherlands  
}

\begin{document}

\maketitle

\begin{abstract}

\BOii\ is one of the most promising systems to determine the Hubble
constant from time-delays in gravitational lenses. Consisting of two bright
images, which are well resolved in VLBI observations, plus one of
the most richly structured Einstein rings, it potentially provides better
constraints for the mass model than most other systems. The main problem left
until now was the very poorly determined position of the lensing galaxy.

After presenting detailed results from classical lens modelling, we apply our 
improved  version of the
\lc\ algorithm which for the first time utilizes the beautiful Einstein ring
for lens modelling purposes. The primary result using isothermal lens models
is a now very well defined lens position of 
$(255\pm6,119\pm4)\,$mas relative to the A image,
which allows the first reliable measurement of the Hubble
constant from the time-delay of this system. The result of
$H_0=(78\pm6)\,\kmsmpc\ (2\,\sigma)$
is very high compared with other lenses. It is, however,
compatible with local estimates from the \hst\ key project and with WMAP
results, but less prone to systematic  errors.
We furthermore discuss possible changes of these results for different radial
mass profiles and find that the final values cannot be very different from
the isothermal expectations.
The power-law exponent of the potential is constrained by VLBI data of the
compact images and the inner jet to be $\beta=1.04\pm0.02$,
which confirms that the mass distribution is approximately isothermal
(corresponding to $\beta=1$), but 
slightly shallower. The effect on $H_0$ is reduced from the expected 4~per
cent decrease by an estimate shift of the best galaxy position of ca.~4\,mas
to at most 2 per cent.

Maps of the unlensed source plane produced from the best \lc\ brightness
model show a typical jet structure and allow us to 
identify the parts which are distorted by the lens to produce the radio ring.
We also present a composite map which for the first time shows the rich
structure of \BOii\ on scales ranging from milli-arcseconds to arcseconds,
both in the image plane and in the reconstructed source plane.
Finally we use a comparison of observations at different frequencies to
investigate the question of possible weakening of one of the images by
propagation effects and/or source shifts with
frequency. The data clearly favour the model of significant `extinction'
without 
noticeable source position shifts.

The technical details of our variant of the \lc\ method are presented in the
accompanying \papi. 

\end{abstract}


\section{Introduction}

The lensed B0218+357 \citep{patnaik93} is one of the rapidly growing but still
small number of gravitational lens 
systems with an accurately known time-delay \citep{biggs99,cohen00}. 
It is therefore a candidate to apply Refsdal's method to determine
the Hubble constant $H_0$ \citep{refsdal64b}.
Besides the time-delay and an at least partial knowledge of the other
cosmological parameters, only good mass models of the lens are needed to
accomplish this task. No other complicated and potentially
incompletely known astrophysics enters the calculation or contributes
to the errors. Since the cosmological parameters are believed to be known
with sufficient accuracy now, e.g.\ from the WMAP project \citep{spergel03},
the only significant source of errors lies in the mass models themselves.
The difficulties in constraining the mass distribution of lenses
should not be underestimated, but the lens method is still much
simpler to apply than classical distance-ladder methods, where
errors difficult to estimate can enter at each of the numerous steps.

The classical methods still fight with the problem of
results which are incompatible within the formal error bars.
The HST key project \citep{mould00,freedman01} e.g.\ obtains a value of
$H_0=(71\pm6)\,\kmsmpc$ ($1\,\sigma$) which is often used as a reference. It
should not be forgotten, however, that other groups obtained results which are
significantly different, even though the general distance ladder used is very
similar. \citet{sandage99} determines a value of $53\pm7$
($1\,\sigma$), \citet{parodi00} obtain $59\pm6$ (90\,\%). The
differences in the analyses are various, ranging from the correction of
selection effects and the Cepheid period-luminosity--relation to instrumental
effects. The fact that a number of world-expert groups spent such an amount of
effort but still do not agree on the final result proves how difficult the
problems of distance ladder methods really are. See the references above for a
discussion of possible reasons for the discrepancies.
In addition we refer to \citet{tammann01} for an extensive discussion of this
subject and an additional result of  $60\pm5$.

After finishing a first version of this paper, the analysis of the
results of the first year of WMAP observations have been
published. Although $H_0$ is not the primary goal of CMB observations,
determinations of this parameter are possible with additional
assumptions. Assuming an exactly spatially flat universe
($\Omega\sub{tot}=1$) and a cosmological constant for the `dark
energy' (equation of state parameter $w=-1$), \citet{spergel03} obtain
a value of $72\pm5$. For different $w$, the result for $H_0$ can be
significantly smaller.  The assumptions of $w=1$ and $\Omega\sub{tot}=1$
can only be dropped if additional information from other astronomical
fields are included. With the local information from the 2dF Galaxy
Redshift Survey \citep{percival01,verde02}, a constraint on $w$ is
obtained. WMAP alone can constrain $\Omega\sub{tot}$ only weakly,
leading to wide allowed ranges for the density of dark energy or
matter. For a determination of all free parameters, more astronomical
data has to be included as discussed in detail by \citet{spergel03}.
The constraints for all cosmological parameters obtained from the
combination of many data sets are very impressive, but it should be
kept in mind that the situation now becomes similar to the distance
ladder methods where model assumptions in many astrophysical fields
are necessary, making realistic error estimates of the final result
extremely difficult. Therefore the more parameters that go into CMB
models have accurate independent determination, the better. This allows the
CMB analyses to concentrate on other parameters which cannot be determined in
any other way.

Taking these difficulties into account, we believe that the lens
method is superior to most other 
methods not only in terms of the possible accuracy of $H_0$ but also,
and not less important, in terms of estimating the possible errors
reliably. We emphasize that the effort needed is small when compared with
large projects like the HST key project or WMAP plus 2dF survey etc.

The only serious problem when using the lens effect to determine the
Hubble constant lies in the mass models for the lenses which directly
affect the resulting $H_0$.  To be able to constrain a mass model
accurately, two conditions should be met. First, the lens should be
simple by which we mean the galaxy should be normal without nearby companions
and without a 
surrounding cluster. Only in these cases can the mass distribution be
described with confidence by a small number of parameters. Second, as many
observational constraints as possible should be provided by the
system. Advantageous is a high number of highly structured
images. Extended lensed sources and Einstein rings can provide
constraints for a wide range of positions and are therefore preferred
to point-like multiple images. In the case of more than two images,
multiple time-delays can also be used as constraints for the mass
model.  In addition to these pure lensing constraints, additional
information from stellar dynamics or surface photometry of the lens
can be included in the modelling.

Radio loud lensed sources have the potential to provide a whole set
of independent observational constraints additional to those provided
by optical observations.  Flux ratios of the images are less
affected by microlensing and extinction at radio wavelengths than
optical ones. Radio observations also have the advantage that they can
reach much higher levels of resolution than in the optical. Much
smaller substructures in the images can therefore be detected and used
for the lens models.

The lens \BOii, discovered in the JVAS survey \citep{patnaik93}, meets
most of the criteria for a `Golden Lens' which 
allows an accurate determination of $H_0$. The lens is an isolated
single galaxy, and no field galaxies nearby contribute to the lensing
potential significantly \citep{lehar00}.
The lens shows only two images of the core of the source, but
substructure of the images can be used as further constraint. The
main bonus is the presence of an Einstein ring which shows a lot of
substructure in detailed radio maps \citep{biggs01}.
This paper deals with utilizing the information from the ring to constrain the
mass models and to determine $H_0$.

\BOii\ has only one disadvantage: It is the system with the smallest
image separation known, which makes useful direct measurements of
the lensing galaxy's position relative to the lensed images extremely
difficult. Since this parameter is of fundamental importance for the
determination of $H_0$, no reliable results have been possible so
far. \citet{biggs99} present a galaxy position which may be close to
the correct value, but their error bars are highly underestimated as
was already stated by \citet{lehar00}.

In this paper we will first discuss `classical' lens models which only use
parametrized data of the compact images to constrain the models. We will see
that with this approach the constraints are not sufficient to determine the
galaxy position with any
accuracy. To be able to determine the Hubble constant, other information has
to be included which is naturally provided by the highly structured radio
ring. The best method available to utilize this information is \lc\
\citep{kochanek92}. Because the method in its original form has serious
shortcomings which prevent its use in a system like \BOii, where the dynamic
range between the compact images and the ring is very high, we had to improve
the algorithm considerably. The \lc\ method itself is discussed in
\papi\ \citep{paper1} while the results are presented here (\papii).

We will use the improved version of \lc\ to determine the galaxy position for
isothermal models with
an accuracy that is sufficient to achieve a result for the Hubble constant
which is competitive with other methods but avoids their possible systematic
errors. Possible deviations from isothermal mass distributions, which are the
main source of systematic error when combining the results from many
lenses, 
will be discussed by presenting a preliminary analysis of VLBI observations of
the substructure of the images. These will allow us to constrain the radial mass
profile for power-law models.
We will learn that
deviations from isothermal mass distributions do \emph{not} play a major role
in \BOii, because the constraints are already quite strong and in addition the
effect on $H_0$ is in this special case, where the galaxy position is
determined indirectly with \lc, much weaker than in most other systems.
We nevertheless discuss future work which will improve the results even more.

As a secondary result we will use our best lens models and newly developed
methods to produce maps of the source plane as it would look like without the
action of the lens. We will show the source as well as its appearance in the
lensed image plane on scales from milli-arcseconds to arcseconds.

Finally we use \lc\ to shed some light on the question of scattering induced
effective `extinction' in the A component of \BOii\ by comparing results for
data sets of different frequencies.

This \papii\ as well as \papi\ are condensed versions of major parts of
\citet{phd}. For more details the reader is referred to that work.

\begin{figure*}
\includegraphics[width=0.7\textwidth]{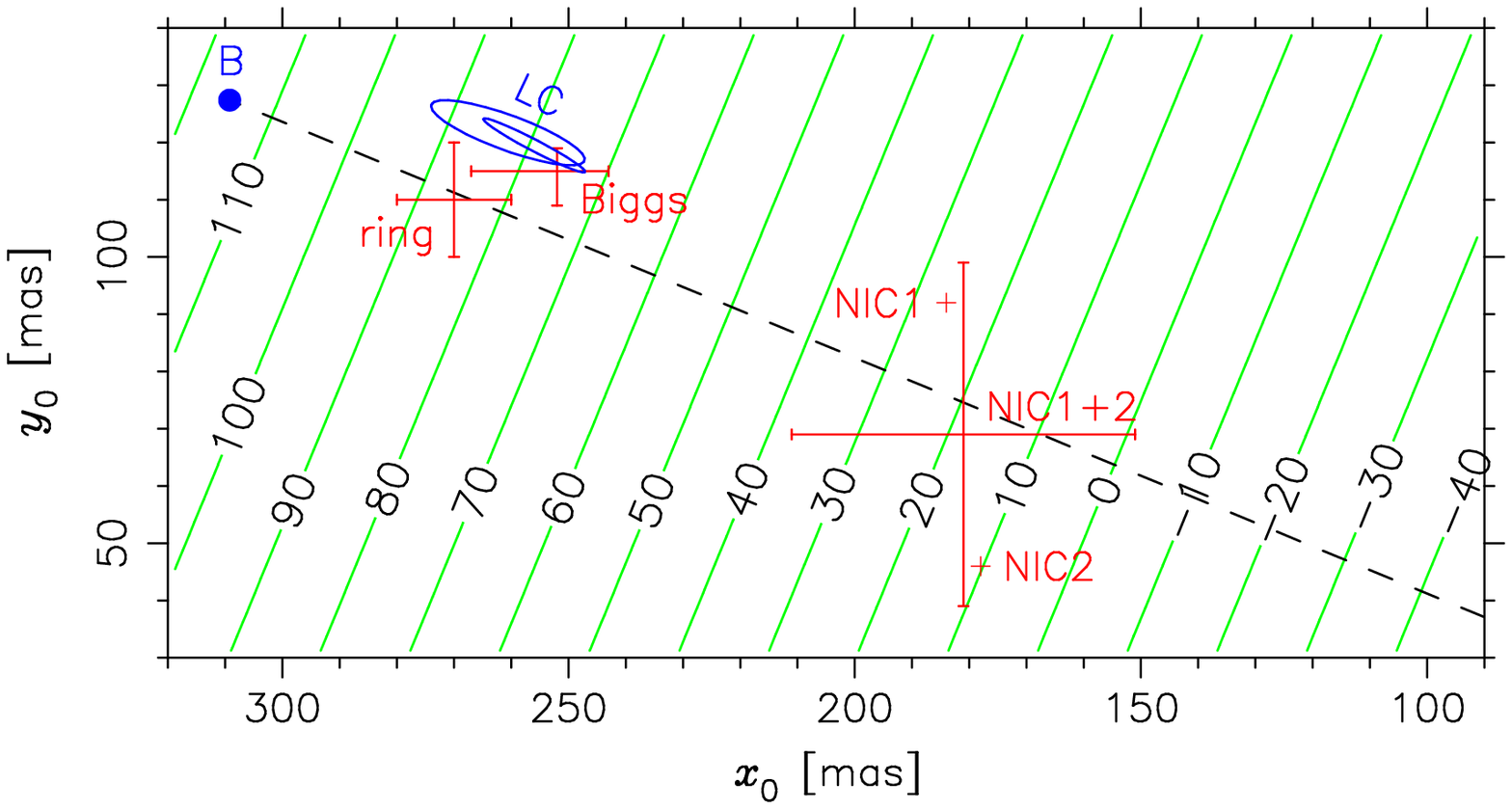}
\caption[$H_0$  as a function of galaxy position]{$H_0$ in units of
  $\rmn{km\,s^{-1}\,Mpc^{-1}}$ as a function  
  of galaxy position $z_0$ (for the concordance model, $\Omega=0.3$,
  $\lambda=0.3$).
  The dashed line connects components A (at the origin) and B.
 Shown are also different  
  estimates for $\vc z_0$. NIC1 and NIC2 are optical positions from
  \citet{lehar00}. At the upper left are the results from \citet{biggs99}
  as well as the centre of the ring \citep{patnaik93}.
  Our final results with \lc\ are shown by the ellipses (LC) to
  the upper left of the `Biggs' error cross. These ellipses mark the formal
  $2\,\sigma$ uncertainties for the two parameters $x_0$ and $y_0$ with
  natural (smaller) and uniform (larger ellipse)
  weighting from Fig.~\ref{fig:vla I}. The $2\,\sigma$ region for the one
  parameter $H_0$ would be slightly smaller.
}
\label{fig:h0}
\end{figure*}

\section{Mass models}

In the absence of an accurate position of the lens relative to the
position of the lensed images there is no
need to use very complicated and general lens models. The effect of
the lens position shows its full impact already with simple standard
models. We therefore restrict most of our work to isothermal
elliptical mass models.  To introduce the notation, we write the lens
equation as
\begin{equation}
\vc\zs = \vc z - \vc\alpha(\vc z) \rtext{.}
\label{eq:lens}
\end{equation}
The true source position $\vc\zs$ is the image position $\vc z$
shifted by the apparent deflection angle $\vc\alpha$. In static single-plane
lenses, the latter can
be written as gradient of a potential $\psi$:
\begin{equation}
\vc\alpha (\vc z) = \vc\nabla\psi (\vc z)
\nt
\end{equation}
We use this potential to parametrize the mass models. General
power-law models have
\begin{equation}
\psi (\vc z) = r^\beta \,F(\theta) \rtext{.}
\label{eq:power law pot}
\end{equation}
The radial power-law index is $\beta=1$ in the isothermal case.
The arbitrary azimuthal function $F(\theta)$ describes the
asymmetry. Elliptical \emph{mass distributions} can be described by the
formalism presented in \citet{kormann94} or \citet{kassiola93}. We
prefer the approximation of elliptical \emph{potentials}
\citep{kassiola93}, which have practical advantages over the true
elliptical mass distributions, because the inversion of the lens
equation is possible analytically for elliptical potentials. For
sufficiently small ellipticities, both approaches are equivalent.
A singular elliptical power-law potential (`SIEP' in the isothermal case) is
given by 
\begin{gather}
\psi(\vc z) = \frac{\alpha_0^{2-\beta}}{\beta} \, \re (\vc z) \nt
\intertext{with}
\re^2 = \frac{x^2}{(1+\epsilon)^2} + \frac{y^2}{(1-\epsilon)^2} \rtext{.} \nt
\end{gather}
For small ellipticities $\epsilon$ and moderate $\beta$, this model
corresponds to an 
elliptical mass distribution with axial ratio $q=1-2\epsilon(2+\beta)/\beta$,
or $q=1-6\,\epsilon$ for isothermal models.
We do not include external perturbations, because the external shear
and convergence
introduced by field galaxies and large scale structure are expected to
be very small. \citet{lehar00} estimate them to be of the order
$0.01$ to $0.02$, which is negligible in our context.

The time-delay $\Delta t_{i,j}$ between images $i$ and $j$ is
generally related to the potential and the apparent deflection angle
in the following way: 
\begin{align}
\Delta t_{i,j} &= \frac{1+\zd}{H_0}\frac{\dd\,\ds}{\dds} \Delta T_{i,j} \nt \\
T_i &= \frac12 |\vc\alpha(\vc z_i)|^2-\psi(\vc z_i) \nt
\end{align}
Here $\zd$ is the redshift of the lens and $\dd$, $\ds$ and $\dds$ are
the normalized angular size distances to the lens, source and from lens to
source, respectively.

Isothermal models have an interesting analytical property, first
described by \citet{witt00}. For fixed positions of the images
relative to the mass centre, the time-delay for given $H_0$ is
uniquely determined and 
does not depend on any other parameters of the model, as long as the
images are fitted exactly. In our case this means that, given the
time-delay, $H_0$ for isothermal models can be calculated directly from the
lens position $\vc z_0$ using the equation
\begin{equation}
T_i = \frac12 |\vc z_i - \vc z_0|^2 \rtext{.}
\nt
\end{equation}
The dependence of $\Delta T$ on $\vc z_0$ is linear.

These simple relations show the importance of measuring the lens position
directly. Once
this position is known with sufficient accuracy, no explicit model fits are
necessary, but the Hubble constant can be derived directly from the positions
of the images and the lens.

The resulting Hubble constant for isothermal models using the time-delay of
$\Delta t= 10\fd5\pm 0\fd4$ \citep{biggs99} is shown in  
Fig.~\ref{fig:h0} as a function of $\vc z_0$. These values are calculated for
a  low-density flat universe ($\Omega=0.3$, $\lambda=0.7$). They would be
smaller by 5.5~per cent for an Einstein-de Sitter universe (EdS).
We use a lens redshift of $z\sub d=0.6847$
(\citealp{browne93,carilli93}; and others)
and a source 
redshift of $z\sub s=0.944$ \citep{cohen02}. The older value of $z\sub s=0.96$
\citep{lawrence96} would lead to a 4.2~per~cent smaller Hubble constant.
Included in the plot are estimates for the lens
position from different sources. The optical positions from
\citet{lehar00} are clearly not accurate enough. The centre of the
ring is relatively well defined, but it can deviate from the position of
the mass centre by 0\farcs03 or even more, depending on the ellipticity of
the lens and the size and shape of the source.
We will see later that the error bars of \citet{biggs99} are underestimated
significantly so that their result for the lens position can not be used.

The alternative time-delay of $10\fd1\pm 1\fd5$ ($2\,\sigma$) from
\citet{cohen00} is compatible with the result from \citet{biggs99} which we
used for our work because of the smaller error bars.
The result for $H_0$ obtained from the \citet{cohen00} time-delay would be
4~per cent higher.

\section{Classical isothermal lens models}

Our first modelling attempts for \BOii\ followed the classical route
of using only parameters of the two compact
components as constraints. This
is the same approach as presented in \citet{biggs99}. We will learn
that it is not possible to determine uniquely the lens position with this
information. 
The results shown in this section were obtained using isothermal models because
the (small, see below) deviations from isothermality are not of importance as
long as the galaxy position is not known with high accuracy. Once this
parameter is determined with \lc, constraints on these deviations as well as
the impact on $H_0$ will be discussed in Sec.~\ref{sec:non-iso}.

\subsection{Observational data}

All coordinates in this paper are measured with
respect to the A image. The $x$ coordinate is measured eastwards
(increasing right ascension), $y$ is measured northwards.

Table~\ref{tab:pos} shows a compilation of known relative positions of
the A and B images as well as of the subcomponents 1 and 2
which were revealed by VLBA observations \citep{patnaik95}.
There are significant differences in the given positions, but they are
too small to affect the results seriously. For the B$-$A separation, we 
used the 15\,GHz VLBA positions for the core components A1 and B1 with their
formal accuracy.
The relative positions of the \emph{sub}components offer an additional
possibility to 
constrain the lens models. For the modelling we assumed an accuracy of
0.1\,mas as in \citet{biggs99} and made independent calculations for the
data from \citet{patnaik95} and \citet{kemball01}. The difference between the
two sets of positions is of the same order as the assumed error bar.

The new VLBI data presented in \citet{biggs02} show, in addition to the
already known two close subcomponents, the inner jet in both images with
subcomponents at separations of up to ca.~10\,mas which could potentially
provide additional constraints. These data are not
used for our classical modelling because the accuracy of parametrized model
fits to the additional components are influenced significantly by the
underlying smooth emission of the jet. The true errors of such fits are
expected to be much higher than formal estimates so that we decided to exploit
this new data set only later with \lc.

\begin{table}
\begin{minipage}{8cm}
  \caption{Relative positions of components and subcomponents.}
\label{tab:pos}
      \begin{tabular}{rl@{~}cd{3.2}d{3.2}c}
        \multicolumn{3}{c}{data set} &
        \multicolumn{1}{c}{$x~[\mas]$} & 
        \multicolumn{1}{c}{$y~[\mas]$} & (sub)comp \\ \hline
        5\,GHz  & EVN & \footnote{\footstrut from \citet{patnaik93}, given to
        $0.6\,\mas$ accuracy}
        & 308.5   & 130.3 & B$-$A   \\
        8.4\,GHz& VLBI & \footnote{\footstrut from \citet{kemball01}, accuracy
        0.09\,\mas\ in each coordinate}%\savefoot
        & 309.00 & 127.30 & B1$-$A1    \\
        & &
        & 309.32 & 126.37 & B2$-$A2    \\
        15\,GHz & VLBA & \footnote{\footstrut from \citet{patnaik95}, given to
        $0.6\,\mas$ accuracy} 
        & 309.2   & 127.4 & B1$-$A1    \\
        &      &  & 309.6   & 126.6 & B2$-$A2 \\
        15\,GHz & VLA & \footnote{\footstrut own fit (uniform weighting), same
        data as for \lc} 
        & 310.56  & 127.11 &  B$-$A \\
        optical & HST & \footnote{\footstrut from \citet{lehar00}}
        & 307     & 126    &  B$-$A \\ \hline
        8.4\,GHz & VLBI & \footnote{\footstrut from \citet{kemball01},
        accuracy ca.~0.05\,\mas\ in each coordinate}
        &    1.18  &   0.87  & A2$-$A1 \\
        &&&  1.50  & -0.06 & B2$-$B1 \\
        15\,GHz  & VLBA & \footnote{\footstrut from \citet{patnaik95},
        accuracy not given}
        &    1.072 &   0.868  & A2$-$A1 \\
        &&&  1.470 &   0.000  & B2$-$B1
      \end{tabular}
    \end{minipage}
\end{table}

The magnification ratio of the two compact images is more
difficult to obtain. Several effects can influence the results. One
problem is scattering in the lensing galaxy which seems to decrease
the flux density of 
A. This effect is stronger for lower frequencies and (with regard to the
fluxes) thought to be 
almost absent at 15\,GHz. Secondly, flux densities of compact
images embedded in a smooth surface brightness
background cannot be determined unambiguously. We always have to
expect an uncertainty of about the background surface brightness integrated
over the beam. This
error should be smaller than 10 per cent for the lower resolution data
sets at lower frequencies and much less than that in other cases.
Finally we observe the source at different epochs in the two images
because of the time-delay. From measured light curves and the known
time-delay \citep{biggs99}, the typical error of this effect is estimated to
be less than 5 per cent.

Table~\ref{tab:flux} shows flux density ratios for different
frequencies, epochs and resolutions.
There seems to be a systematic increase in the ratio with increasing frequency
(discounting optical), which hints at scattering. Also the spread in values for
8.4\,GHz and above is relatively small. For both these reasons, and because
there are values with the effects of delay removed, we chose to use a value of
3.75 which is compatible with most of the measurements at higher
frequencies. Please 
note that the results depend only very weakly on small changes of this
ratio. For the \lc\ work presented below, no explicit measurement of the flux
density ratio is required. Resolution effects and contamination by the ring
are then taken into account automatically.

\begin{table}
\caption[Flux ratios for different frequencies]{Flux density
  ratio A$/$B for different frequencies and instruments.}
\label{tab:flux}
\begin{minipage}{8cm}
\renewcommand{\thefootnote}{\mbox{$\alph{footnote}$}}
  \begin{tabular}{d{1}@{\,}ll@{~}ccd{1.3}@{}l}
    \multicolumn{3}{c}{data set} && epoch & \multicolumn{2}{c}{ratio} \\ \hline
    1.7&GHz& VLBI & \footnote[1]{\footstrut from \citet{patnaik99}}
     & 1992 Jun 19 &
    2.62 \\
    5&GHz & MERLIN& \footnote[2]{\footstrut from
    \citet{patnaik93}}
    & 1991 Aug 26 & 2.976 \\
    5&GHz & MERLIN& \footnotemark[2]
    & 1992 Jan 13 & 3.23 \\
    5&GHz & MERLIN& \footnotemark[2]
    & 1992 Mar 27 & 3.35 \\
    5&GHz & EVN & \footnotemark[2]
    & 1990 Nov 19 & 3.185 \\
    8.4&GHz & VLA& \footnotemark[2]  & 1991 Aug 1 & 3.247 \\
    8.4&GHz & VLA& \footnote[3]{\footstrut from \citet{biggs99}, simultaneously fitted
      with time-delay, used for their models}  & 1996/1997   & 3.57 &
    $\pm 0.01$ \\
    8.4&GHz & VLA& \footnote[4]{\footstrut from \citet{cohen00}, varying part
      simultaneously fitted with time-delay}
    & 1996/1997   & 3.2 & $\pm0.35$ \\
    8.4&GHz & VLBI& \footnote[5]{\footstrut subcomponents A1/B1 \citep{kemball01}}
    & 1995 May 9 & 3.18 & $\pm0.17$ \\
     &  &  & \footnote[6]{\footstrut subcomponents A2/B2 \citep{kemball01}}
    &  & 3.72 & $\pm0.20$ \\
    15&GHz & VLA& \footnotemark[2]   & 1991 Aug 1 & 3.690 \\
    15&GHz & VLA &\footnote[7]{\footstrut own fit (uniform weighting), same
      data as for \lc}   & 1992 Nov 18 & 3.79 \\
    15&GHz & VLA&\footnote[8]{\footstrut from \citet{biggs99}, simultaneously fitted
      with time-delay}   & 1996/1997   & 3.73 &  $\pm 0.01$ \\
    15&GHz & VLA& \footnotemark[4]  & 1996/1997  & 4.3 & $\pm0.65$ \\
    15&GHz & VLBA &\footnote[9]{\footstrut from \citet{patnaik95}} 
    & 1994 Oct 3 &
    3.623 & $\pm 0.065$ \\
    22&GHz & VLA&\footnotemark[2]   & 1991 Aug 1 & 3.636 \\
    \multicolumn{2}{c}{5550\,\AA} & HST & \footnote[10]{from \citep{castles02,lehar00}}
 && 0.14 &
  \end{tabular}
 \end{minipage}
\end{table}

The subcomponents are only marginally
resolved and even the formal uncertainties of fitted Gaussians
are too high to serve as additional constraints. Furthermore, the
apparent sizes and shapes might be influenced by scatter broadening in the
interstellar medium of the lensing galaxy, which is known to be gas
rich \citep{wiklind95}. We nevertheless tried to include the shapes of the
subcomponents 
into the modelling, without any gain in accuracy of $H_0$.
The actual observational data as given by \citet{patnaik95} and
\citet{kemball01} are shown in
Table~\ref{tab:subcomp shapes}. The fit of the best source shapes for
each lens model is performed analytically with a Cartesian parametrization of
the ellipses. Mathematical details can be found in Appendix~\ref{sec:app
  shapes}. 

\begin{table}
  \caption[Shapes of the subcomponents]{Shapes of the
    subcomponents}
  \label{tab:subcomp shapes}
\begin{minipage}{8cm}
\renewcommand{\thefootnote}{\mbox{$\alph{footnote}$}}
\begin{tabular}{ccccr@{ $\pm$}r}
    & & major axis $[\mas]$ & minor axis $[\mas]$ &
    \multicolumn{2}{c}{p.a. $[\deg]$} \\ \hline
    A1& \footnote[1]{15\,GHz \citep{patnaik95}}
                 & $0.58\pm0.05$ & $0.28\pm0.05$ & $-37$ &  5 \\
      & \footnote[2]{8.4\,GHz \citep{kemball01}}
                 & $1.90\pm0.30$ & $1.12\pm0.30$ & $157$ & 6 \\[1ex]
    A2& \footnotemark[1]  & $1.02\pm0.05$ & $0.54\pm0.05$ & $-47$ &  5 \\
      & \footnotemark[2]  & $\ge2.0$      & $\ge2.0$ \\[1ex]
    B1& \footnotemark[1]  & $0.36\pm0.05$ & $0.16\pm0.05$ & $-65$ & 15 \\
      & \footnotemark[2]  & $0.72\pm0.30$ & $0.35\pm0.04$ & $73$  & 12 \\[1ex]
    B2& \footnotemark[1]  & $0.61\pm0.05$ & $0.23\pm0.05$ & $83$  & 10 \\
      & \footnotemark[2]  & $\ge0.73$     & $\ge0.73$
  \end{tabular}\end{minipage}
\end{table}

\begin{figure*}
\includegraphics[width=0.9\textwidth]{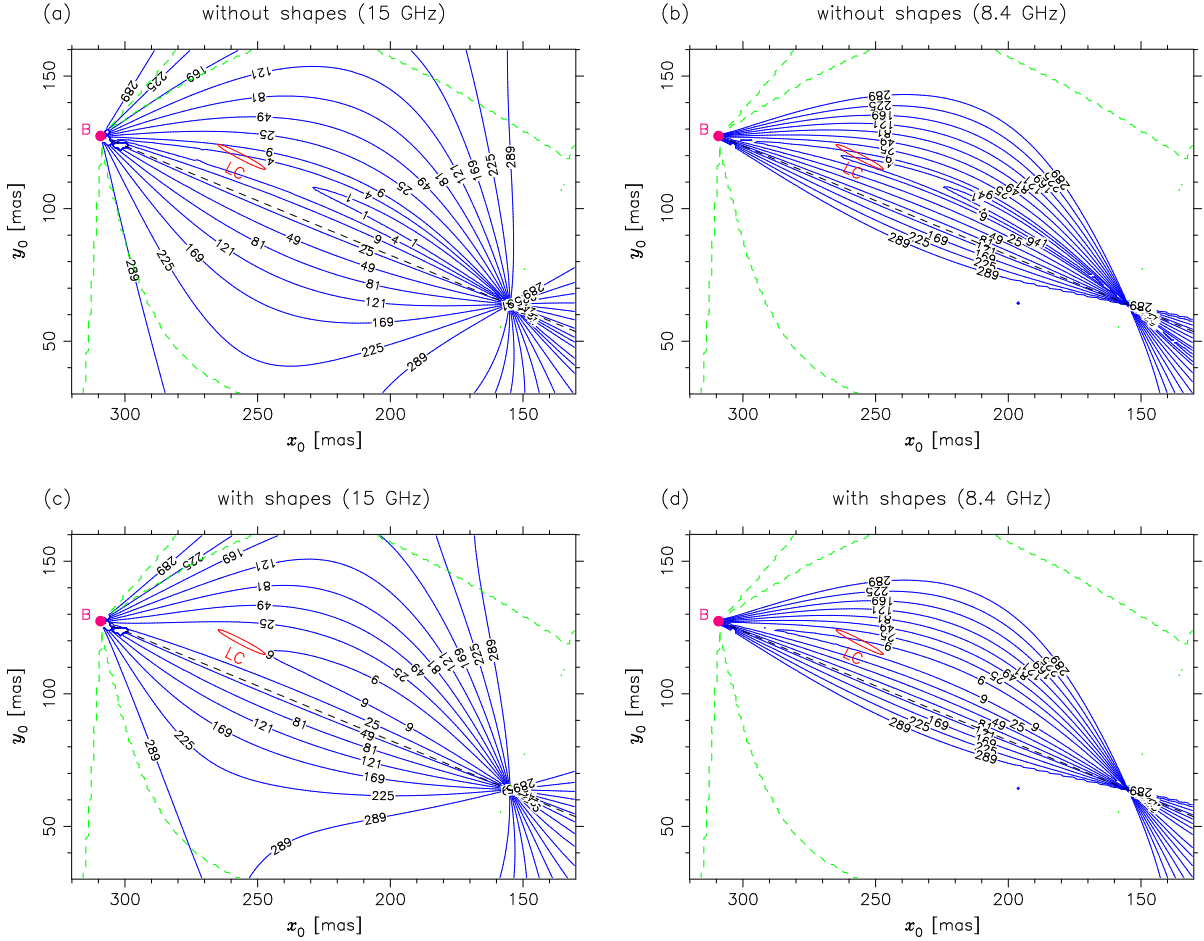}
\caption[$\chi^2$  as a function of galaxy position]{Best $\chi^2$ for
  isothermal models as a
  function of galaxy position. Subcomponent positions and shapes for the left
  plots are from
  15\,GHz VLBA data \citep{patnaik95}, for the right plots from
  8.4\,GHz VLBI data \citep{kemball01}. 
 Constraints for the models are the
  relative position of A and B, the flux density ratio and the
  positions of the VLBA subcomponents.
The upper plots are without, the lower ones with the
  subcomponent shapes included as constraints.
  The dashed curves separate allowed regions with two images in the
  central parts of the plots from regions where more than two images
  would be seen.
The small ellipse marked LC is the $2\,\sigma$ confidence region of our final
  \lc\ result for natural weighting, see Fig.~\ref{fig:vla I}\,(b).
}
\label{fig:chi2}
\end{figure*}

\subsection{Results}

Before proceeding with numerical models, we want to compare the numbers of
 parameters and constraints to learn what can be achieved in the
absence of degeneracies.
The SIEP models introduced above have a total of five free
parameters (mass scale $\alpha_0$, position $x_0$, $y_0$ and ellipticity
 $\epsilon_x$, $\epsilon_y$), not counting the source position. The positions
 of A and B 
give us two constraints (or four constraints minus two parameters for
the true source position), the flux density ratio another one. The
relative positions of the subcomponents provide another two totalling
in five constraints. These numbers may seem promising but we have to
be aware of possible degeneracies which might prevent us from
determining all parameters and $H_0$. This is in fact the case.

Without using the subcomponents, only three constraints are
available. From this small number of
constraints, it is impossible to determine the parameters of our mass
model. It is in fact possible to find exactly one fitting model for each
given galaxy position $\vc z_0$ plotted in Fig.~\ref{fig:h0}. A wide
range of (positive and formally negative) values of $H_0$ is compatible
with these data.

In the next step, we include relative positions of the
subcomponents from Table~\ref{tab:pos}. See Appendix~\ref{sec:app vec} for
details on how these data are used in the modelling.
Naively, we would expect these
two more numbers to constrain the galaxy position significantly.
Figure~\ref{fig:chi2}\,(a)+(b) shows the total $\chi^2$ of the best fitting
models for a range of fixed galaxy positions. For these models, we used
a flux density ratio of 3.75, but other values measured at high frequency
would lead to similar results. The relative parity of
the two images was fixed to be negative as it is necessary for double image
lens models. With our fitting algorithm, this was already sufficient to
exclude models with more than two images in those regions where double lenses
are at all possible.

As the plots show, the subcomponent positions provide only one more
effective constraint, confirming the result of \citet{lehar00}.
This is easily explained. In the best models shown,
the subcomponents are located nearly exactly radially with respect to
the mass centre. Since singular isothermal models show no dependence of
the radial stretching on image positions (there is no radial
stretching at all), we lose one constraint. The only constraint left
acts perpendicular to the line connecting A and B. This is
unfortunately not the direction in which $H_0$ changes (see
Fig.~\ref{fig:h0}).
For galaxy positions leading to realistic ranges of $H_0$ (like our \lc\
results which will be presented later), the minimal $\chi^2$ is about 1.6 for
the 15\,GHz VLBA data \citep{patnaik95} and 3.1 for the 8.4\,GHz VLBI data
from \citet{kemball01}. The formal number of degrees of freedom is $\nu=0$ in
this case, but the model degeneracies lead to an effective value of
$\nu=1$.\footnote{If the radial mass index $\beta$ is allowed to vary freely,
  the lowest residuals come very close to zero.} 
The fact that the data can be fitted quite well is
interesting, because it shows that the data are at least approximately
compatible with isothermal models.

We already noted that the shapes of the subcomponents are not a
very promising candidate for further constraints.
Figure~\ref{fig:chi2}\,(c)+(d) confirms this view.
The $\chi^2$ contribution from the shapes alone is not sensitive for shifts in
the radial direction and the total $\chi^2$ function does not change
qualitatively. 
The relatively high $\chi^2$ is a hint that the measured shapes are not
exactly the correct ones, just as one might expect if the shapes are affected
by scattering. The ineffectiveness of shape constraints does not change
when non-isothermal models are used. This is a result of the fact that most
information about the relative magnification matrix is already provided by the
magnification ratio and the radial stretching of the subcomponent separations.

We conclude that the available data on the two images are
compatible with lens positions leading to an extremely wide range of
values of $H_0$ and we can therefore not confirm the relatively small
error bars given by \citet{biggs99}, even if (in contrast to
\citealp{lehar00}) the subcomponent shapes are included as
constraints.
Without any further information about the position of the lensing
galaxy, `classical' lens modelling using only the compact components
does not provide any useful estimates for $H_0$.\footnote{{\hst}/ACS
  observations 
  performed to measure the galaxy position directly are currently being
  analysed. They will provide a value for this parameter which might be
  formally less 
  accurate than our \lc\ results but which is completely
  independent of the lens model. (PI: N.~Jackson)}

\section{LensClean}

\subsection{The method}

One promising way to get more information about the mass distribution of
the lens is to make use of the extended emission which shows as a ring
in radio maps \citep[e.g.][]{biggs01}. Several approaches for models using
extended emission have been discussed in the literature
\citep[e.g.][]{kochanek89,kochanek92,ellithorpe96,wallington96}. They all work
by, for a given lens model, constructing a map of the source 
which minimizes the deviations from observations when mapped back to
the lens plane. The minimal residuals themselves are then used in an
outer loop to find the best lens model.

In \papi\ we describe the \lc\ method which was first proposed by
\citet{kochanek92} and \citet{ellithorpe96} together with a number of
improvements and the details of our own implementation.
The idea of \lc\ as an extension of standard \clean\ is very simple. 
\clean\ builds up a radio map by successively collecting point sources at the
peaks of the so-called dirty map, subtracting them from the observed data
and in the end convolving the collection of components with a Gaussian in
order to produce the \clean\ map. \lc\ works similarly, but 
for each position in the
source plane, emission has to be subtracted at the positions of \emph{all}
corresponding image positions simultaneously with the magnifications
given by the (for now fixed) lens model. In this way an emission model is
built which is exactly compatible with a given lens model and which comprises
the best fit to the data under this assumption.
An outer loop to determine the lens model parameters themselves is then built
around the \lc\ core. In this loop the remaining residuals of \lc\ are
minimized by varying the lens model parameters. The result is a combined
best-fitting model for the lens mass distribution and the brightness
distribution of the source.

For our work we used a data set taken at the VLA in 1992 at 15\,GHz with a
bandwidth of 50\,MHz and a total on-source time of about 6~hours.
The beam size is $129\times146\,\mas^2$ (p.a. $-73\,\degr$) with natural
and $86\times 88 \,\mas^2$ (p.a. $-60\,\degr$) with uniform weighting.
The thermal rms noise per beam is 55\,$\umu$Jy for natural and
140\,$\umu$Jy for uniform weighting.
The same data were used in the development of \lc\ described in \papi. More
details can be found there.

The standard pixel and field size used for the final runs is a $512\times512$
pixels of $5\times5\mas^2$ each. A loop gain of $g=0.38$ is used and 5000
iterations of our improved unbiased \lc\ variant are performed. 
Smaller pixels and a smaller loop gain would make the residual function
smoother, which helps in finding the minimum, but would increase the computation
times on the other hand. Systematic deviations caused by
too large pixels or too high $g$ are not significant for the parameter ranges
used. 
For the first ungridded step which removes most of the bright compact images,
a loop gain of $\gamma=0.95$ is used. As magnification limit for the images
we chose a value of 100.

As explained in more detail in \papi, we scanned the range of possible lens 
positions $\vc z_0$ and fitted the remaining lens model parameters for each
position in order to stabilize the fits and to be able to determine confidence
regions for the position from the final maps of residuals as a function of
$\vc z_0$.
To reduce the remaining
numerical noise, smooth polynomial functions are fitted to the residual
function $R^2(\vc z_0)$ to determine the minimum and confidence regions.

Possible changes of the flux ratio A$/$B as a result of the time-delay in
 combination with intrinsic variability are taken into account with the
methods described in \papi.
From the residuals of the best models
we learned that any ratio changes have to be less than 5~per~cent. This is in
good agreement with monitoring data covering the epoch of our observations
which showed an almost constant flux ratio of $3.79\pm0.07$ during that time
\citep{corbett96}.
The measured flux ratio is therefore a good estimate of the true magnification
ratio, almost unaffected by the combination of variability and time-delay.
Taken together, these results also confirm that, even if scattering or other
propagation effects 
in the interstellar medium (ISM) of the lensing galaxy may change the flux
ratio at lower 
frequencies, the effect is not relevant at 15\,GHz and can thus be neglected
here. We will later estimate the effect at 5\,GHz by comparing with our
results for 15\,GHz.
The shifts
of the best $\vc z_0$ caused by flux ratio changes below 5~per~cent are not
significant and will be neglected. 

Because of the numerical difficulties of the \lc\ algorithm in combination
with lens models which do not allow an analytic inversion of the lens
equation, most of our calculations are done for isothermal (SIEP) models
without external shear. As explained above and in \papi, these relatively
simple models are 
already sufficient to show the effect of the very important lens position on
the determination of $H_0$. We will discuss later that the true radial mass
profile seems to be very close to isothermal and we will estimate the effect
of this deviation as a perturbation of the SIEP model.

\begin{figure*}
\includegraphics[scale=1.6]{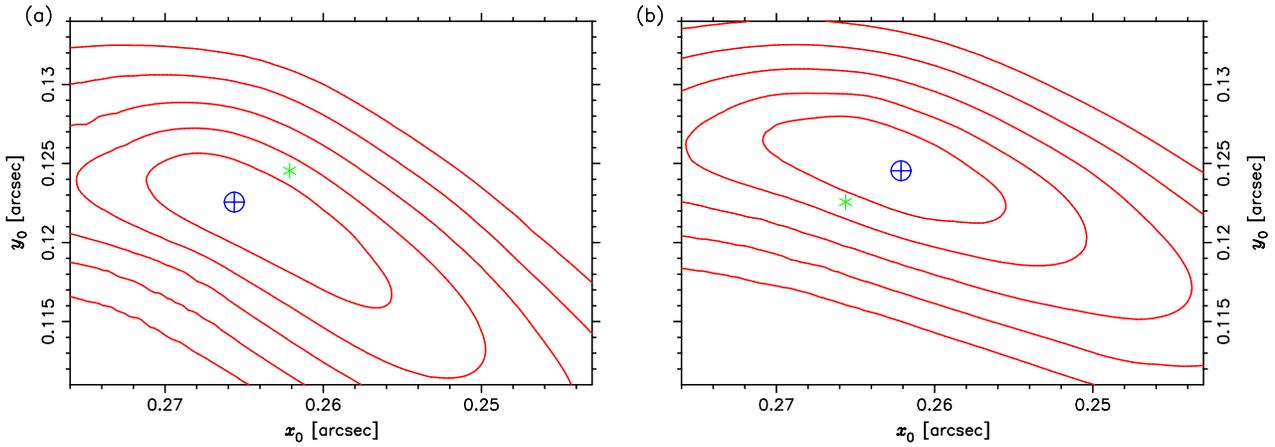}
\caption[\lc\ residuals 15\,GHz VLA LL and RR]{Final residuals 
for the 15\,GHz VLA data set for (a) LL and (b) RR polarization (uniform
 weighting) for isothermal models. The  results 
 are smoothed for the
 contour lines (confidence limits of 1, 2, 3, 4, 5$ \cdot \sigma$).
The cross hair marks the respective residual minimum, the asterisks the one
 of the other polarization.}
\label{fig:vla LL+RR}
\end{figure*}

\begin{figure*}
\includegraphics[scale=1.6]{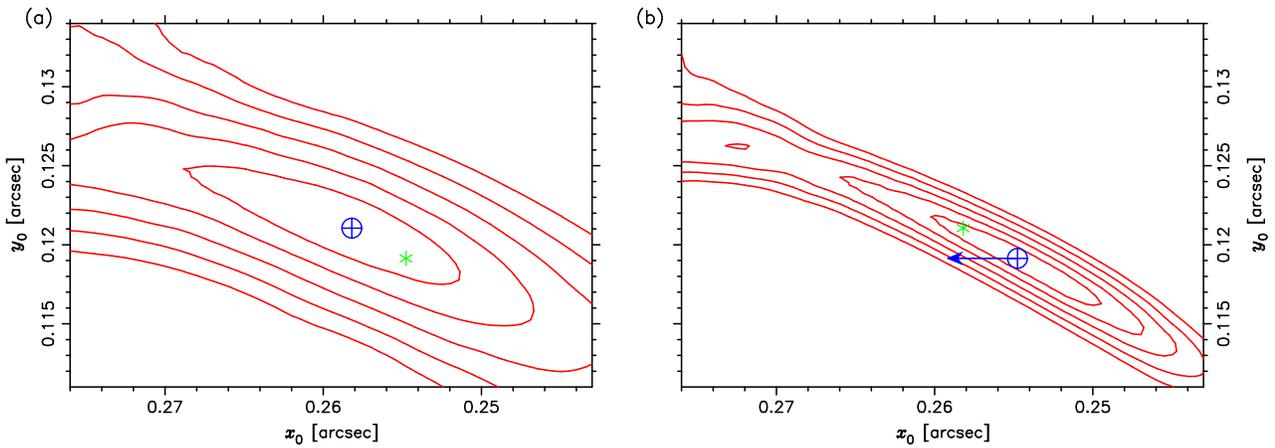}
\caption[\lc\ residuals 15\,GHz VLA I]{Final residuals 
for the 15\,GHz VLA data set for Stokes I for isothermal models. The
 results are
 smoothed for the
 contour lines (confidence limits of 1, 2, 3, 4, 5$\cdot \sigma$).
 (a) uniform, (b) natural weighting. The cross hair marks the
 respective residual minima, the asterisk marks the minimum for the
 other weighting scheme.
 The arrow in (b) symbolizes the estimated shift for a
 non-isothermal model with $\beta=1.04$ (see Sec.~\ref{sec:noniso lc}). The
 shape of the residual function would be very similar for such a model.}
\label{fig:vla I}
\end{figure*}

\subsection{Determining the lens position}

We started the analysis with the left and right handed circular polarization
separately to be able to use the comparison of the two as consistency check.
\lc\ with self-calibration as described in \papi\ was used independently on the
uniformly weighted LL and RR data set to produce final $uv$ data sets and
residual maps. The residuals as a function of the galaxy position (relative to
A) are shown in Fig.~\ref{fig:vla LL+RR}. To produce the plot, the residual
function was interpolated and smoothed from the original results which were on
a relatively coarse grid consisting of ca.\ 200 points.

The confidence regions are estimated
from Eq.~\eqref{I-eq:DR2} in \papi\ and the $\chi^2$ distribution as described
there. 
The best normalized residuals are $0.77$ and $0.80$ for LL and RR, while
the expectation from Eqs.~\eqref{I-eq:R2} and \eqref{I-eq:sigma2} in \papi\
would be $1\pm 
0.009$. Our values are significantly lower because of the
self-calibration applied to the data. We allowed independent phase and
magnitude correction for each integration bin independently which, by the high
number of free parameters, naturally reduces the residuals considerably.
\clean{}ing and self-calibrating the data with \difmap\ without taking into
account the lens leads to the very similar residuals of 0.78. We conclude
that the best fits with \lc\ have residuals within the expectations but have
to keep in mind that this is not a very sensitive test for the goodness of
fit. The best galaxy positions from LL and RR are clearly compatible within
$1\,\sigma$.

Some care is necessary when combining the two circular polarizations to obtain
total intensity I which is done in the next step. As a starting lens model for
self-calibration we used the 
mean of the best galaxy positions for LL and RR, obtained the other parameters
for this position separately for LL and RR and took the mean of the two. This
provides a better estimate than taking the mean of the best models of LL and
RR directly. The \lc\ calculations were done as before but in the
self-calibration possible differences in the calibration errors of LL and RR
had to be taken into account. To do this we built the emission model for I
(mean of LL and RR) and calibrated LL and RR separately with this emission
model. Afterwards the two were again combined to obtain total intensity I.
This standard approach relies on the assumption that no circular polarization
is 
present and that LL and RR should (apart from noise) be the same.
With uniform weighting only positive components were allowed for
self-calibration. For natural weighting, we started with 3000 uniformly
weighted iterations, switched to natural weighting and performed another 2000
iterations. This is superior to 5000 iterations with natural weighting because
the improved resolution of the uniformly weighted beam helps in avoiding
positional errors of the first bright components. Because of this weight
switching, negative components were allowed as well. They are essential
to compensate for the difference of uniform and natural weighting.

The results are shown in Fig.~\ref{fig:vla I}.
The plot does not show the result
from the fits directly but was produced by first fitting the models for a
relatively coarse grid
of $\vc z_0$ with \lc, then fitting smooth polynomials to the resulting model
parameters and calculating \lc\ residuals on a finer grid where the polynomial
fits where used for the lens model parameters. This approach does not change
the appearance of the plot but saves a large amount of CPU time because real
\lc\ fits have to be performed only for a much smaller number of galaxy
positions than needed to produce a smooth plot.

The final error ellipses 
are also included in Fig.~\ref{fig:h0} and \ref{fig:chi2}.
Although uniform weighting was superior to natural weighting in the first
stages of calibrating the data, in the end the naturally weighted results are
slightly more accurate if we trust the formal statistical uncertainties.
The accuracy depends on two aspects here. One is the resolution of the maps,
which is better for uniform weighting, the other is the signal to noise ratio,
which is optimal for natural weighting. The tradeoff between the two aspects
depends on the lens models and on the data set and it is not a priori clear
which weighting scheme is optimal for \lc.

We see that the \lc\ lens position is located almost exactly in the centre of
the valley of low residuals of classical lens model fitting
data shown in 
Fig.~\ref{fig:chi2}. The results are therefore consistent with each
other. Since the two methods use very different constraints (the VLBI
substructure in classical and the ring in \lc\ modelling), this agreement
provides good evidence that the \lc\ methods actually works correctly and
produces reliable results.

The statistical uncertainties in the total intensity (I) result are
only slightly smaller than in the separate LL and RR results. Slightly
unexpected, the best lens position from I (uniformly weighted) is not located
between the best 
positions from LL and RR but has an offset of ca.\ 6\,mas, still compatible
within the $1\,\sigma$ bounds.
This effect is not fully understood yet. It might be related to low-level
circular polarization in the source which would demand a different
self-calibration technique. More probable is the presence of subtle
instrumental effects which could not be corrected for by self-calibration.

\begin{figure*}
\includegraphics[width=0.9\textwidth]{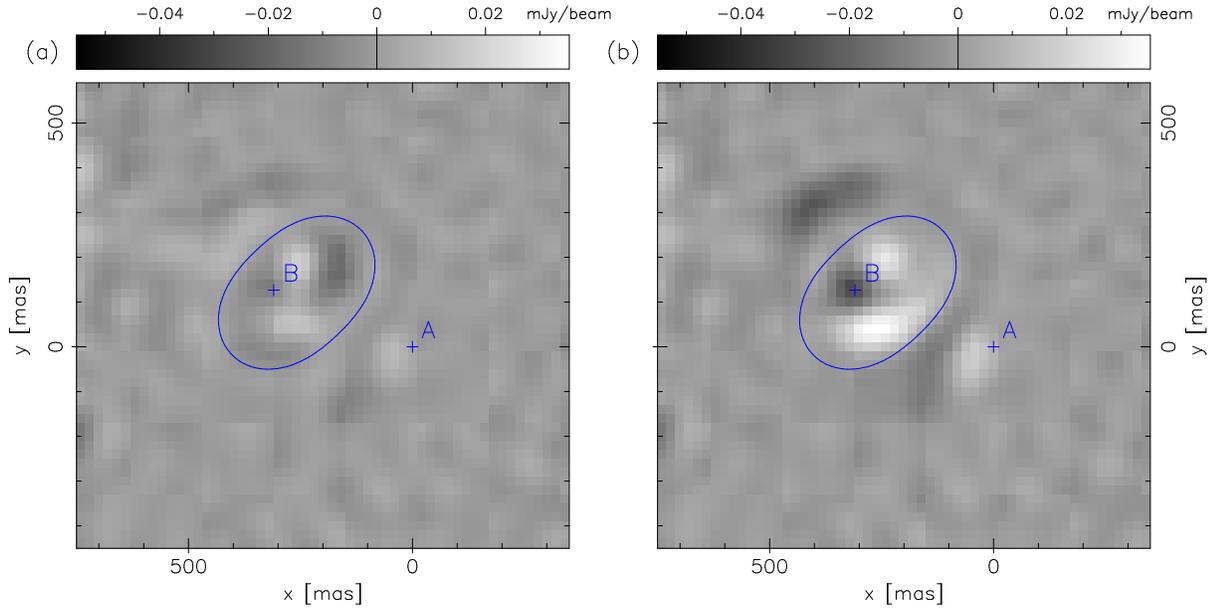}
\caption[Image plane residuals]{Uniformly weighted image plane
  residuals after 100\,000 iterations of \lc\ with
  (a) the best lens model, (b) a lens model that is excluded by $1\,\sigma$.
  The expected noise of the dirty maps is 0.14\,mJy per beam. Compare these
  maps with their simulated equivalents in Fig.~\ref{I-fig:resid1}\,(c+d) of
  \papi. 
  The critical curve and the two bright images are marked.
}
\label{fig:resid2}
\end{figure*}

The final residuals for the naturally weighted data set and the best lens
model are reduced $\chi^2=0.842$. This is again significantly below unity as a
result of self-calibration. The value is compatible with results using
simulated data sets but should not be used as test for the goodness of the
fit. For the latter purpose we show in Fig.~\ref{fig:resid2} the uniformly
weighted (to increase the resolution) image space
residuals for the best lens model (from uniformly weighted data) and an
alternative model that is off by $1\,\sigma$ (calculated for one parameter).
For these maps we used a very high number of 100\,000 iterations, many more
than the 5000 used for the model fits. We notice that in the outer singly
imaged regions, where the \clean{}ing is not constrained by the lens, most of
the noise has been included in the models so that the residuals are much
smaller than the rms noise of the data. For the best model shown in
Fig.~\ref{fig:resid2}\,(a) even the inner 
parts, which really depend on the lens model, are \clean{}ed down to a similar
level. This proves that the lens model can not be too far from the correct
one. For the alternative model, on the other hand, the inner residuals are
significantly higher than before, see Fig.~\ref{fig:resid2}\,(b).
These residual maps should be compared with Fig.~\ref{I-fig:resid1}\,(c+d) in
\papi\ which shows the same for a simulated data set using a known lens
model. The residuals of our alternative model are of the same level as in the
simulations. For the best-fitting model, the real residuals are even smaller
which could be surprising at the first view but which results from the fact
that the best-fitting lens model has \emph{by definition} the smallest
residuals which are thus always smaller than residuals for the correct lens
model.
We conclude that the total $uv$-space residuals and the image space residual
map are both consistent with a perfect fit of the lens model.

In order to test our estimates of confidence regions in the case of uniform
weighting, we performed Monte Carlo simulations the results of which are shown
in Fig.~\ref{I-fig:mc} in \papi. The scatter of simulated results is in good
agreement with the 
theoretical expectations although these could only be obtained by using
approximations for the expected statistical properties of the residuals. In
the case of natural weighting, the residuals follow the standard $\chi^2$
distribution.

\subsection{Best isothermal lens model and Hubble constant}

The best fitting lens models are given in Table~\ref{tab:lensmod}. The errors
are mainly due to the uncertainty in $\vc z_0$. For a fixed $\vc z_0$ they
would be smaller by orders of magnitude.
The real flux ratio calculated from the models is $3.80\pm0.03$ which is
compatible with the uniformly weighted direct model fitting to the same data
set. The agreement with the value of 3.73 resulting from the time-delay
analysis in \citet{biggs99} is also very good.

The ellipticity of the mass model is unexpectedly large. We have to remember
that the 
ellipticity of an equivalent elliptical mass distribution is three times the
ellipticity of the potential which leads to a best estimate of the axial ratio
of $a/b\approx 1.4$.
In contrast to this, the lensing galaxy appears quite round on optical images.
The uncertainties in
$\epsilon$ are quite large, however, so the real ellipticity might be
smaller.
For ellipticities as high as observed here, it is of concern how far the
elliptical potential still is a good approximation for elliptical mass
distributions. As was shown by \citet{witt00}, the magnification in
self-similar isothermal models is a direct function of the surface mass
density which means that the critical curve shown in Fig.~\ref{fig:lens map}
is also an isodensity contour. The shape of this
curve looks rather elliptical and does not show hints
of the unrealistic dumbbell shapes which are observed for more extreme
ellipticities. We conclude that the use of elliptical potentials is still
justified in our case.

\begin{table}
\caption[Parameters of the best fitting lens models]{Parameters of the best
  fitting isothermal lens models for uniform and natural 
  weighting (final result VLA 15\,GHz Stokes I). The error bars are
  statistical $2\,\sigma$ limits for one parameter each; they originate mostly
  from the uncertainty of the lens position and would be much smaller for
  fixed $\vc z_0$. The position angle of
  the major axis $\theta$ is measured in the usual astronomical sense from
  north through east.}
\label{tab:lensmod}
\begin{tabular}{c@{}cll}
parameter    &  & \multicolumn{1}{c}{uniform} & \multicolumn{1}{c}{natural} \\ \hline
\vphantom{$\displaystyle\frac12$}
$ x_0 $ & $[\text{arcsec}]$ &
$\phantom{+}0.2582 \pam{+0.0158}{-0.0087}$ & $\phantom{+}0.2548 \pam{+0.0062}{-0.0063}$ \\
\vphantom{$\displaystyle\frac12$}
$ y_0 $ & $[\text{arcsec}]$ &
$\phantom{+}0.1210 \pam{+0.0057}{-0.0048}$ & $\phantom{+}0.1191 \pam{+0.0031}{-0.0039}$ \\
\vphantom{$\displaystyle\frac12$}
$ \alpha_0 $ & $[\text{arcsec}]$ &
$\phantom{+}0.1616 \pam{+0.0039}{-0.0124}$ & $\phantom{+}0.1634 \pam{+0.0026}{-0.0034}$ \\
\vphantom{$\displaystyle\frac12$}
$\epsilon_x$ & &
$+0.0043 \pam{+0.0123}{-0.0438}$ & $+0.0091 \pam{+0.0066}{-0.0098}$ \\
\vphantom{$\displaystyle\frac12$}
$\epsilon_y$ & &
 $-0.0707 \pam{+0.0273}{-0.0637}$ & $-0.0585 \pam{+0.0196}{-0.0221}$ \\
\vphantom{$\displaystyle\frac12$}
$|\epsilon|$ & &
$\phantom{+}0.0709$ & $\phantom{+}0.0592$ \\
\vphantom{$\displaystyle\frac12$}
$\theta$ & $[\text{deg}]$ & 
$-46.7$ & $-49.5$ \\
\vphantom{$\displaystyle\frac12$}
$\Delta T$ & $[\text{arcsec}^2]$ &
$\phantom{+}0.0393\pam{+0.0055}{-0.0032}$ & $\phantom{+}0.0380\pam{+0.0023}{-0.0025}$
\end{tabular}
\end{table}

The Hubble constant $H_0$ can now be determined from the distances and from
the parameter $\Delta T$ given by the lens model.
For the final result we only use the naturally weighted data since the
uncertainties are smaller here.
For an EdS world model, the result is 
$H_0=(74.1\pm5.5)\,\kmsmpc$.
For a
low-density flat 
standard model with $\Omega=0.3$ and $\lambda=0.7$, it increases by about
6~per~cent:\footnote{The result from the uniformly
  weighted data would be larger by 3.4 per cent 
which is consistent with the naturally weighted result within $1\,\sigma$.}
\begin{equation}
H_0=(78.5\pm5.8)\,\kmsmpc
\nt
\end{equation}
The error bars are formal $2\,\sigma$ limits which include the uncertainty of
the lens model (6 per cent, contributed almost exclusively by the uncertainty
of the lens position) and the time-delay (4 per cent).
We stress that they do not include possible
systematic errors which might result if the true mass distribution can not be
described by our isothermal models (SIEP). As we show below, no large effects
from this are expected.
Using the time-delay from \citet{cohen00}, our result becomes
$H_0=(82\pm13)\,\kmsmpc$ with the largest error contribution (15 per cent)
coming from the time-delay.

Standard quintessence models \citep[e.g.][]{linder88b,linder88a} have
parameters between the 
EdS and the low-density $\Lambda$ world models and would lead
to Hubble constants between the two extremes. For $w=-1/3$, the result would
be indistinguishable from the EdS case (equivalent to $w=0$) while for
$w=-2/3$ the Hubble constant would be 2.1~per~cent higher than for EdS. The
cosmological constant corresponds to $w=-1$.

\section{Non-isothermal lens models}
\label{sec:non-iso}

It is a well known fact the the radial mass distribution is the most important
source of uncertainties in the $H_0$--$\Delta t$ relation for lens systems
where the lens position is known accurately
\citep[see e.g. references in the introduction of][]{kochanek02}. 
For power-law models a simple scaling 
law of $H_0\propto 2-\beta$ is valid in many cases
\citep{wambsganss94,witt95,wucknitz99} 
while in quadruple lenses an even stronger dependence of $H_0\propto
(2-\beta)/\beta$ was found under certain circumstances by \citet{wucknitz02}
for generalized power-law models following Eq.~\eqref{eq:power law pot}.
As a generalization of the SIEP models, we use elliptical power-law potentials
which (if not too far from isothermal and not too elliptical) are a good
approximation to elliptical power-law mass distributions.
The two families of models differ only in the special form of the azimuthal
function $F(\theta)$.

We will see that deviations from isothermality are very small so that our \lc\
results for isothermal models can still be used as a first approximation. The
deviations can thus be treated as a small perturbation.

\subsection{Constraints from classical modelling}

The subcomponents of both images are oriented more or less radially with
respect to the expected lens position. This has the disadvantage that their
positions cannot be used to constrain the galaxy position for isothermal
models as shown before. For non-isothermal models, on the other hand, this
insensitivity can 
be used to determine constraints for the radial mass profile from these data
relatively independent of the true lens position. If this position is taken to
be close 
to the \lc\ result and compatible with the VLBI constraints, $\beta$ can be
estimated with some accuracy using the method presented in
Appendix~\ref{sec:app vec}. 
If we use the 8.4\,GHz VLBI data from \citet{kemball01}, the result is
$\beta=1.04\pm 0.02$ ($1\,\sigma$ limits). The error bar does not include
uncertainties of the galaxy position but the values are relatively independent
of $\vc z_0$ anyway. From the 15\,GHz data from \citet{patnaik95}, a slightly
larger value of $\beta=1.06\pm0.03$ can be estimated. 
\citet{biggs02} were able to detect the jet with many subcomponents over a
length of 10\,mas in both images with global VLBI observations at 8.4\,GHz. A
systematic relative stretching of the B jet image by about 10~per~cent can be
accounted for by a non-isothermal lens model with $\beta\approx1.04$, in very
good agreement with the other results.
We conclude that the lensing galaxy seems to be close to isothermal with a
deviation in $\beta$ of only about 4~per~cent, but have to keep in mind that
these results are only preliminary estimates which are not based on a
self-consistent lens model but used the lens position from \lc\ with
isothermal models to estimate deviations from isothermality. The errors of
this method are expected to be only small (see also below) but a combined fit
of non-isothermal lens models to the VLA and VLBI data should be aimed for in
the future.

\subsection{Preliminary results from \lc}
\label{sec:noniso lc}

As explained in \papi, \lc\ relies on a very robust method for the
inversion of the lens equation. Using the newly developed \lentil\ method
implemented into \lc, we 
were now able to perform a few tests with non-isothermal lens
models. We repeated the calculations
described before with the 15\,GHz VLA data but used different fixed values of
$\beta$ and compared the results afterwards.
We noticed that the best lens position does indeed shift slightly eastwards
when increasing $\beta$, by about 1\,\mas\ for 1~per~cent change of
$\beta$. See the arrow in Fig.~\ref{fig:vla I} for an illustration. To our 
relief, this effect acts on $H_0$ in the opposite direction as the scaling with
$2-\beta$, which we expect for fixed galaxy positions, so that the two cancel
almost completely. A rather conservative estimate is that the $2-\beta$
scaling is reduced by at least a factor of two as a result of the shifting
lens centre. This means that for a value of $\beta=1.04$ a change of the
Hubble constant by at most 2~per~cent is expected compared to the
isothermal models. This is much smaller than the statistical uncertainties
of the result for isothermal models and thus not of concern yet.
When \emph{increasing} $\beta$ slightly, starting from the isothermal
$\beta=1$, the resulting shift of the best \lc\ lens position has the tendency
to \emph{decrease} the optimal $\beta$ in classical fits. Combined fits of
both data sets should therefore be possible and lead to very stable results.

In the near future we will use \lc\ and \lentil\ to obtain more accurate
constraints on the radial mass profile from the existing very high-quality
VLBI observations at 8.4\,GHz \citep{biggs02}. In combination with the
existing VLA data which are more sensitive to the galaxy position, a
self-consistent lens model can be found and accurate error estimates will be
possible.

\section{Source and lens plane maps}

\begin{figure*}
\includegraphics[width=0.9\textwidth]{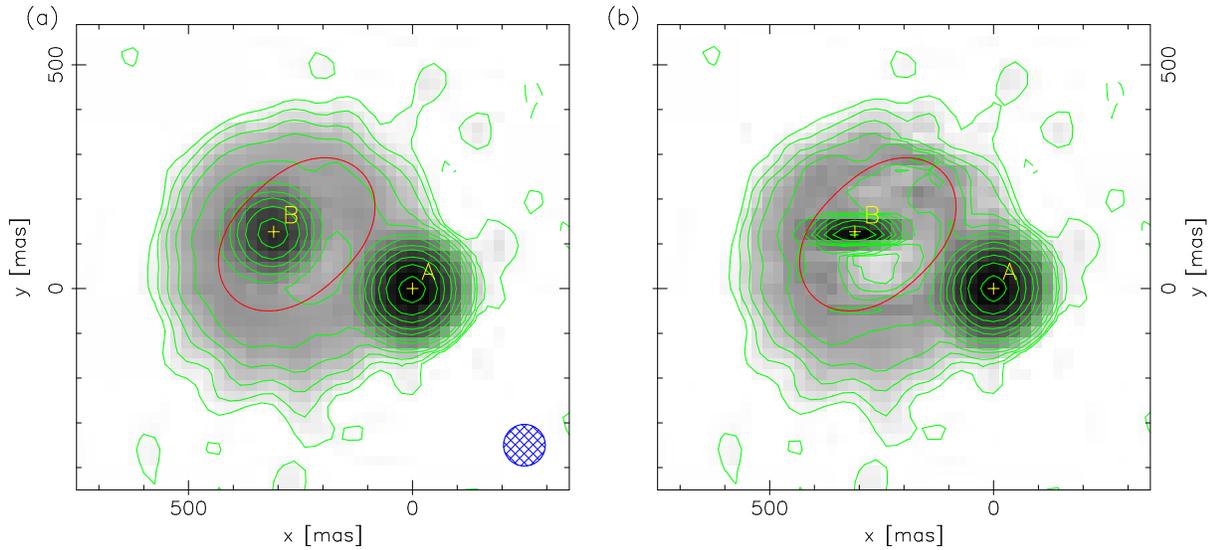}
\caption[Lens plane maps of \BOii]{Lens plane maps of \BOii. (a) Normal \lc\ map at nominal resolution
  of the observations. The $94\times92\,\mas^2$ (p.a. $-57\degr$)
  restoring beam is shown at lower right. (b) Superresolved (see text). The
  lowest contour line in both plots is at 
  $2\,\sigma$ of the expected noise level ($0.14\,\text{mJy}/\text{beam}$
  corresponding to $1.44\cdot10^{-8} \, \text{Jy}/\mas^2$), the value doubles for
  each following line. The critical curve is shown as elliptical line.}
\label{fig:lens map}
\end{figure*}

\begin{figure*}
\includegraphics[width=0.9\textwidth]{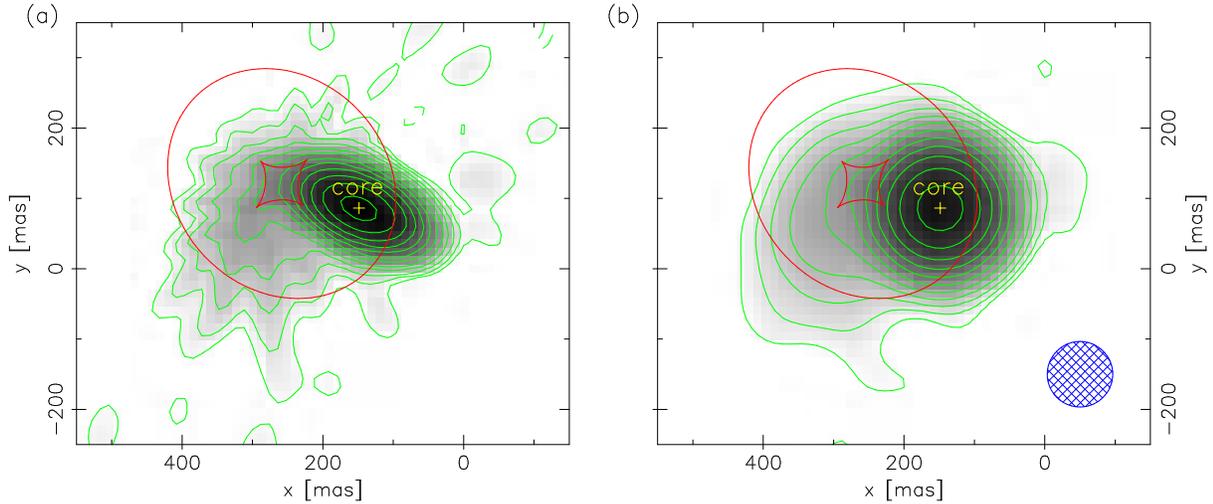}
\caption[Source plane maps of \BOii]{Source plane maps of \BOii. (a) Restored with the anisotropic
  approximated source plane \clean\ beam (see text). (b) Restored with
  circular beams (similar to the lens plane beam shown at lower right).
  The contour levels are the same as in Fig.~\ref{fig:lens map}.
Also shown are the diamond shaped caustic and the
  elliptical `pseudo-caustic' or `cut' \citep{kovner87}.}
\label{fig:src map}
\end{figure*}

In \papi\ we discussed a new method to reconstruct source plane maps and
superresolved lens plane maps from emission models determined with \lc.
We present the resulting lens plane maps of \BOii\ in Fig.~\ref{fig:lens
  map}. One 
version (a) is restored with the nominal \clean\ beam while the other (b) is a
superresolved version with anisotropic and varying beams calculated as
explained in \papi. The superresolved map is not easy to interpret because of the
varying and anisotropic beams. The resolution is generally higher close to the
critical 
curve i.e.\ in parts of the ring. The elliptical shape of the B Image is
also a result of the local elliptical beams.

The reconstructed source plane map is shown in Fig.~\ref{fig:src map}. The
left map (a) is reconstructed with the source plane beams calculated from 
Eqs.~\eqref{I-eq:beam source 1} and \eqref{I-eq:beam source 2} in \papi. The
radial streaks
emerging from the region near the caustic are artifacts of the varying and
highly anisotropic beams. The elliptical shape of the core is also a result of
the highly eccentric local beam and should not be misinterpreted as a
resolved elliptical core.
The right map (b) was restored with varying but
circular beams (the smallest circles which still cover the local elliptical
beams completely) to make interpretations easier. Unfortunately the greatest
part of the lensing magnification and improvement of resolution is lost in
this map.

\begin{figure*}
\includegraphics{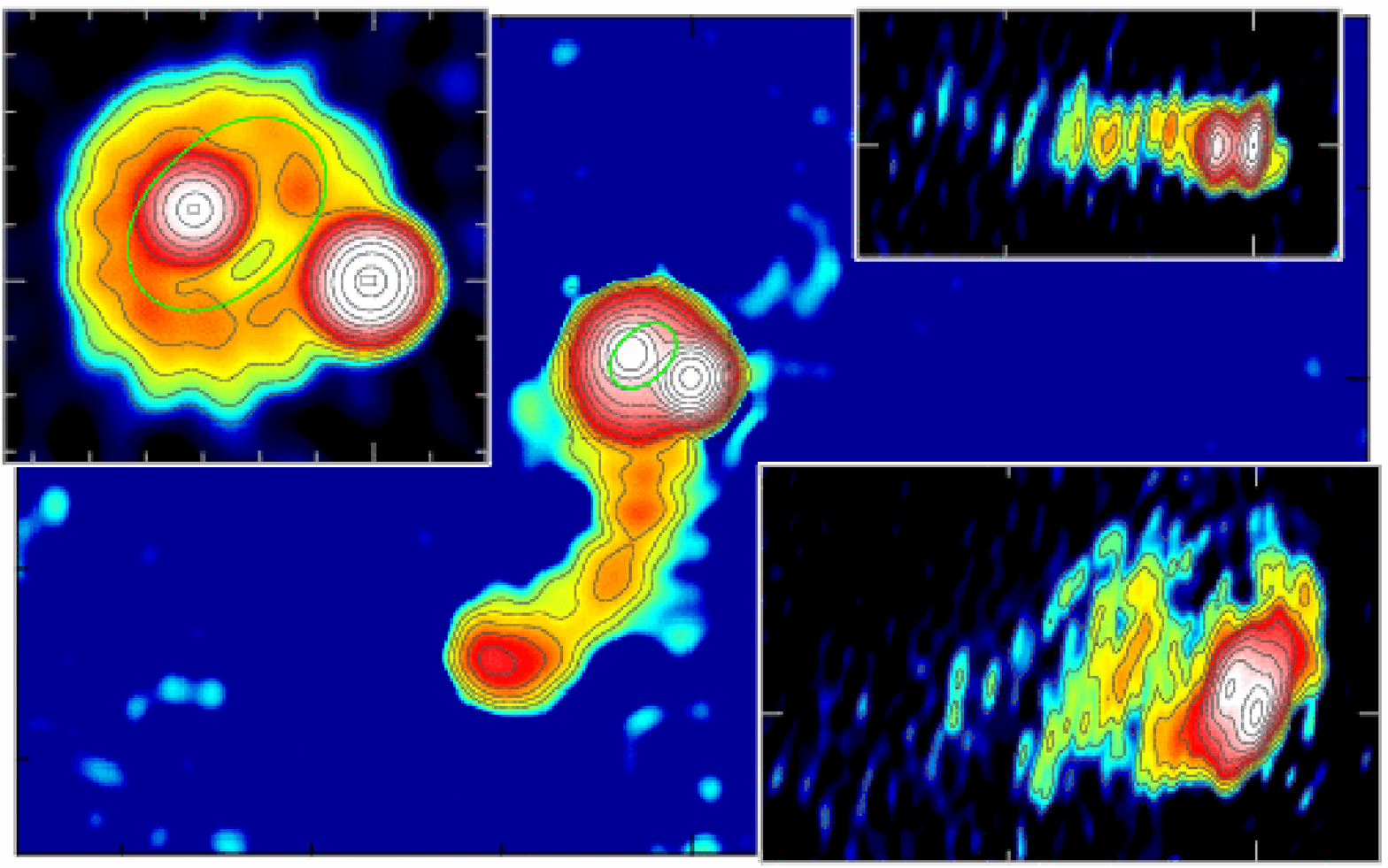}
\includegraphics{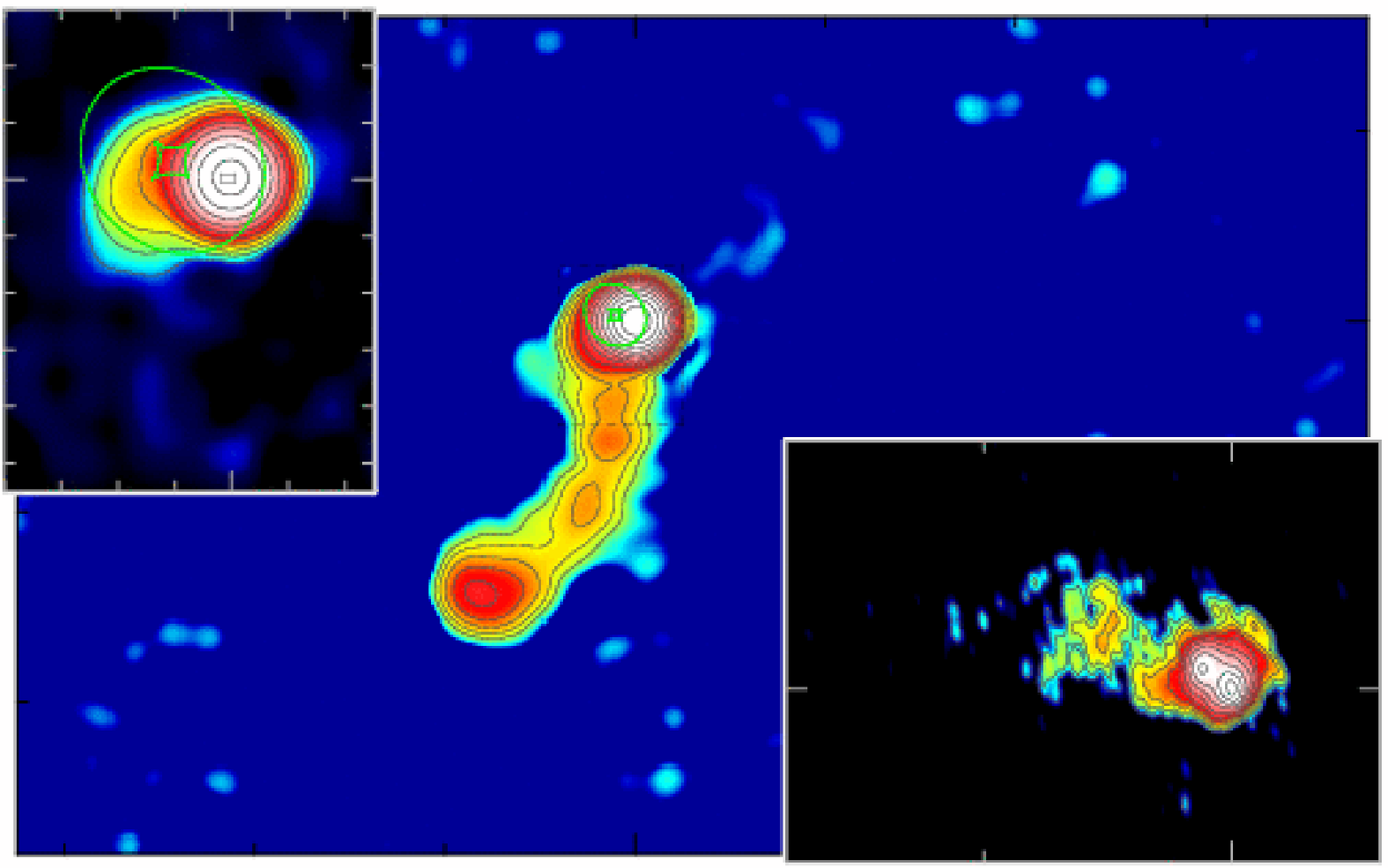}
\caption[Composite image of \BOii]{Composite image of the lens plane (top) and
  source plane  (bottom) of \BOii. The main central images show 8.4\,GHz data
  from the VLA (tick marks 
  1\,arcsec). The insets on the 
  left are from the 15\,GHz VLA data set (tick marks 100\,mas, compare
  Fig.~\ref{fig:lens map} and \ref{fig:src map}), the jets on
  the right 
  are from 8.4\,GHz global VLBI observations (tick marks 10\,mas). The
  magnified regions are marked by black rectangles.
See Section~\ref{sec:comp} on page \pageref{sec:comp} for details.
}
\label{fig:comp}
\end{figure*}

We see that the jet, which by coincidence emerges from the inner core in a
direction very 
similar to the major axis of the elliptical core image in Fig.~\ref{fig:src
  map}\,(a), bends southwards and crosses the caustic of the lens at the
bend. It is this part of the source which forms the ring in the lens plane and
which is used as constraint by \lc. On larger scales (shown especially well by
longer wavelength observations, e.g.\ the 8\,GHz VLA data in
Fig.~\ref{fig:comp}), the jet proceeds in a southern direction only to 
bend eastwards again at a distance of about $1\,\mathrm{arcsec}$. The jet then
seemingly ends in a relatively bright blob $1.5\,\mathrm{arcsec}$ south-east
of the core.

\paragraph*{The composite colour map}
\label{sec:comp}
in Fig.~\ref{fig:comp} shows \BOii\ on different scales
both in the lens plane and in the source plane (reconstructed with \lc).
The main maps showing the outer parts of the jet on arcsec scales are made
from a 8.4\,GHz data set obtained at the VLA simultaneously with a global VLBI
observation \citep[Fig.~4]{biggs02}. We show a regular lens plane map and a
source plane reconstruction with circular beams.
The insets on the left are made from the 15\,GHz VLA data set that was used
for our main \lc\ work. The lens and source plane maps show the same data as
Fig.~\ref{fig:lens map}\,(a) and \ref{fig:src map}\,(b).
We also show the critical curve in the VLA lens plane maps and the tangential
caustic (diamond shape) and cut (elliptical) in the VLA source plane maps.

The insets on the right (upper: B, lower: A) are regular lens plane \clean\
maps of a global VLBI data set \citep{biggs02}. The 
source plane map is a projection into the source plane of the A image
alone. The B image is less resolved and would not provide much additional
information. The map was restored from the \clean\ components with the
projected \clean\ beam. Fig.~5 in \citet{biggs02} shows projections of images
A \emph{and} B restored with equal beams to show the similarities. True \lc\
source plane maps of the VLBI data have not been produced yet.

The magnifications relative to the main maps are 3 for the left insets and 130
for the jet insets on the 
right. The scales are the same for lens plane and
source plane. The contour lines start at three times the rms noise and double
for each new level. The surface brightness levels in the lens plane and source
plane maps are directly comparable.
The magnified sections are marked by black rectangles in the VLA maps.

\section{Investigation of frequency dependent flux ratios}

It is a generally accepted fact that the apparent flux ratio A$/$B changes
systematically with the frequency at which it is measured. At the high end the
value seems to reach its final value of 3.8 for frequencies $\ga
15\,\text{GHz}$. On the other end the ratio goes down to 2.6 at
$1.7\,\text{GHz}$ (compare Table~\ref{tab:flux}). At even lower frequencies it
is difficult to separate the compact emission from the ring which
becomes more dominant at low frequencies.
There is unambiguous evidence that the ISM of the
lensing galaxy is very rich and that optical and radio emission are affected
to a significant degree \citep{wiklind95}. \citet{biggs02} estimated the
scattering measure from 
the broadening of image A relative to B (taking into account the magnifying
effect of the lens). They obtain a value of $\text{SM}\approx 150\,\text{kpc}
\,\text{m}^{-20/3}$ which is very high compared to typical lines of sight along
our galaxy but comparable to lines of sight through the Galactic Centre.

One possible explanation for the low flux ratios at low frequencies is
therefore the action of the very rich ISM of the lensing galaxy in front of
image A. This can cause an effective extinction of A either by a high amount
of scattering or by other physical effects. The refractive index of an
interstellar plasma goes proportional to $\lambda^2$ which would explain the
stronger effects for larger $\lambda$.

Another possible explanation is a very strong frequency dependence of the
source structure which would effectively result in a position shift along the
jet for lower frequencies. Because the spectrum of the jet is much steeper
than the spectrum of the core, such an effect is not implausible. When the
source is shifted in this direction, the situation becomes more symmetric and
the magnification ratio decreases. For a sufficiently large shift ($\ga
10\,\mas$) the 
differential magnification gradient could, in combination with this frequency
dependent source position, well explain the observed flux ratio changes.
There are, however, strong arguments against this scenario. If the shift is as
large as required, there should be significant differences in the appearance
of the innermost components in VLBI maps at 8.4\,GHz
\citep[see e.g.][]{kemball01,biggs02} and 15\,GHz \citep[see
e.g.][]{patnaik95}, but these  
maps look remarkably similar. Both show two strong inner components with
similar shapes and compatible relative positions. Their separation of $\sim
1.4\,\mas$ 
is far too small to explain the demanded shift by different spectral
indices. If the shift is caused by another component which is not seen at
15\,GHz but becomes stronger at low frequencies, such a component should be
detectable on a scale of $10$ to $20\,\mas$ in the existing very deep 8.4\,GHz
VLBI observations. \citet{biggs02} do indeed detect more jet components but
these are far too weak to account for a shift of the required magnitude.
A more exotic alternative scenario would be refraction caused by large scale
systematic trends in the column density of the lens galaxy's ISM.

To investigate the question of extinction and/or source shifts with \lc, we
compare our best lens models fitted to the 15\,GHz data which are known to be
unaffected by both effects with a MERLIN data set at 5\,GHz. This
multi-frequency data set was used to produce the maps in \citet{biggs01} and
is 
described in detail there. We used only the MERLIN part because the VLA
resolution at 5\,GHz is too low to be of any help here.

Unfortunately, this data set has some disadvantages for \lc. While the
resolution is better than in the 15\,GHz VLA data, and the flux in
the ring is higher due to the lower frequency, it was not possible to
obtain reliable direct constraints for the lens model with \lc. One problem
is the possibility of extinction itself.
Another very serious problem is the frequency-dependence of the
emission, which is different for the ring and the compact images. 
\citet{biggs01} approximately corrected for this effect by first
mapping the three frequencies independently. The \clean\ components
responsible for the compact images were then subtracted from the data
so that only the ring remains. The three data sets were rescaled
in magnitude to obtain a consistent total flux density at all
frequencies. Finally, the compact components from the central
frequency were added back to the complete data set. This process may
introduce some distortions to the lowest and highest frequencies,
because it combines data for the ring from these frequencies with data
for the compact components from the central frequency.
The combined data were then successively mapped and self-calibrated to
obtain the final data set and  map.

We used the resulting $uv$ data set
as basis for our computations. The spectral index correction seems to
work very well for making maps, but introduces errors in \lc\ which are
not fully understood. This shows especially in incompatible results from
the individual three frequencies and the combination of all three.

On the other hand, tests showed that the data can nevertheless be used to fit a
subset of the lens model parameters if the other ones are fixed. At most three
parameters can be determined accurately because they are already well defined
by the two bright images alone and the ring is only a relatively small
correction. We therefore used a variety of lens models compatible with the
15\,GHz VLA data and fitted the same models to the 5\,GHz data set allowing
for a relative shift of the two data sets and a possible extinction in the A
image at 5\,GHz. These fits are much more robust than complete lens model
fits from this data set alone. In particular, the results do not
depend on 
which of the frequencies is used. The combination of all three also leads to
the same result.

Our strategy to determine shift and extinction works like this. First we took
one of the best lens models from \lc{}ing the 15\,GHz data. Then we fixed all
parameters but $\vc z_0$ and made a fit with the 5\,GHz data including a (for
the moment fixed) extinction in A. The difference between the two values of
$\vc z_0$ is then a measurement of the shift in the lens plane because in
reality the 
lens should have the same position at both frequencies but the source may
shift. We then varied the extinction to find the best value by minimizing the
residuals.

\begin{figure}
\includegraphics[width=8cm]{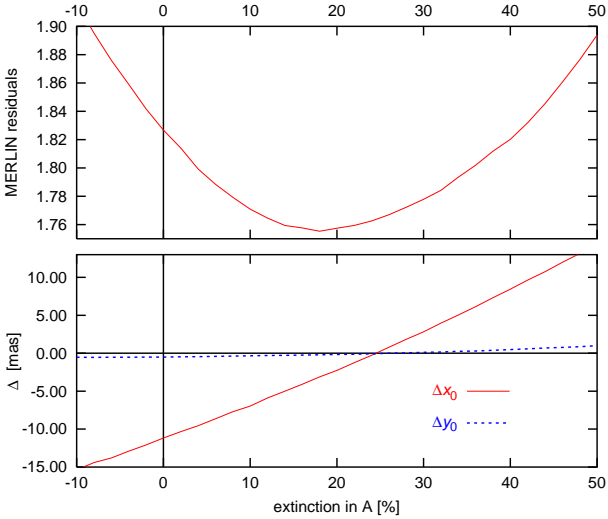}
\caption[Fits of MERLIN 5\,GHz data with models from VLA 5\,GHz]{Fits of
  MERLIN 5\,GHz data with lens models fitting the VLA 15\,GHz 
  data as a function of the extinction in A. top: residuals, bottom: relative
  shift of MERLIN$-$VLA data.}
\label{fig:vlarefit}
\end{figure}

The results for one of the best VLA 15\,GHz lens models are shown in
Fig.~\ref{fig:vlarefit}. The position of the minimum changes somewhat when
other lens models still compatible with the 15\,GHz data are used. The most
extreme values for the extinction are 15 and 35~per~cent; most probable is a
value between 20 and 30~per~cent. The data sets where initially registered to
obtain the same position for the A image. The relative shift in the fit
(measured as a shift of the lens centre MERLIN $-$ VLA, equivalent to a shift
of the source by the same amount with opposite direction) in the region of
the minimum is negligible. The fact that the shifts in $x$ and $y$ have
their zeros at almost the same extinction which itself is furthermore
compatible with the 
residual minimum gives strong evidence that the real shift is indeed small and
that the brightness of A at 5\,GHz is reduced by $\sim 25$~per~cent by
some propagation effect, possibly scattering. For zero extinction, the
required shift would be about $11\,\mas$ eastwards,
but the residual curve seems to be incompatible with this scenario\footnote{We
  do not give a formal significance or confidence limits of extinction or shift
  because the systematic problems with this data set are not fully understood
  yet.}. 
It seems as if extinction in A, caused by whatever physical process, accounts
for a least most of the frequency dependent flux ratio.
This has an important consequence for further modelling work. Although the
ring becomes brighter at lower frequencies, only high frequencies can be used
in a simple way to fit lens models. Observations at medium frequencies may
on the other hand be used to study the propagation effects in the ISM of the
lensing galaxy in more detail now that the lens model is known with sufficient
accuracy.

\section{Discussion}

\BOii\ is one of the most promising lens systems for determining the Hubble
constant with the method of \citet{refsdal64b}. This method has, compared with
other approaches, the advantage that it is a very simple one-step
determination and 
relies on the understanding of only very little astrophysics. In this way the
systematic uncertainties can not only be minimized but (equally important)
also be estimated much better than in distance ladder methods which have their
problems with several astrophysical processes at each step.
The only serious uncertainty left is the mass distribution of the lens. In
this respect \BOii\ is a close to optimal `Golden Lens'. The lensing galaxy
seems to be a regular isolated galaxy without contributions from a group or
cluster nearby. This not only avoids the inclusions of further parameters
to describe the external mass distribution but also allows the assumption that
the mass distribution of the galaxy itself is that of an unperturbed smooth
ellipsoid. 
\BOii\ provides a wealth of potential constraints for the mass models in the
form of highly accurate image positions, a measurable flux density ratio with
probably negligible influence by microlensing, substructure on scales resolved
by VLBI and a well resolved and structured Einstein ring on the scale of the
image separation.

Without using the ring, `classical' lens modelling allows us to tightly
constrain the
ellipticity and mass of the galaxy as long as its position is known. The VLBI
structure of several subcomponents along the jet in a radial direction
relative to the lens allows us to determine the radial mass distribution with
an 
accuracy not possible in most other lenses \citep[see e.g.][and references therein]{kochanek03}. This most important general degeneracy of
the lens method can therefore be broken in the case of \BOii.
The only real disadvantage is the small size ($\sim0\farcs3$) of the
system. \BOii\ is indeed the lens with the smallest image separation known to
date. Although the images and the lens are detected with optical observations,
it 
was until now not possible to use these observations to accurately measure the
galaxy position directly. \hst/ACS observations obtained recently are
currently analysed to allow a first useful direct measurement.

The model lens positions from \citet{biggs99} are probably not too far from
the truth and are indeed consistent with our results but their error bars are
seriously underestimated. We have shown 
that, without using the ring, no useful estimate of the galaxy position and
thus $H_0$ is possible.

Previous efforts have not utilized the most striking feature of
\BOii, 
the beautiful Einstein ring shown by radio observations
\citep[e.g.][]{biggs01}. Doing this is much 
more difficult than using parameters of compact components because the source
itself has to be modelled as well, either implicitly or explicitly. The method
best suited for this task is \lc\ which surprisingly has been used only for
very few cases before. This is partly a result of the high numerical
demands and partly of serious shortcomings of the original method as it was
implemented by \citet{kochanek92} and \citet{ellithorpe96}. In \papi\ we
discuss the development of our new 
improved variant of \lc. One of the most important improvements is the
implementation of a new unbiased selection of components.

The problem of determining the lens position is almost degenerate because the
bright components, which contribute most of the signal, provide no information
for this parameter. It is only the relatively weak and fuzzy ring which can be
used as constraint. In other words the residuals have a very strong dependency
on two directions (defined by the compact components) of the five dimensional
parameter space but change only weakly in the other three directions as a
result of the constraints provided by the ring. Finding minima
of functions like this is a numerically difficult problem and relies on the
very accurate knowledge of the residual function. Any numerical noise produces
local minima which make the finding of the global minimum difficult. Special
care was therefore needed to produce the best possible results with the given
limited computing resources.

The modified \lc\
algorithm was applied to a 15\,GHz VLA data set resulting in good constraints
for all parameters of an \emph{isothermal} lens model, including the lens
position of $x_0=(255\pm6)\,$mas, $y_0=(119\pm4)\,$mas relative to the A image.
This model was then used to determine the Hubble constant from the
time-delay of \citet{biggs99} to be
$H_0=(78\pm6)\,\kmsmpc$. The error bars are $2\,\sigma$
confidence limits which for the Hubble constant include the error of the lens
position (6 per cent) and the time delay (4 per cent). The accuracy for all
other lens model parameters is much higher as long as the position is fixed.
The value for $H_0$ is in agreement with the results from the \hst\ key project
\citep{mould00,freedman01} and the WMAP project \citep{spergel03} but
incompatible with the lower distance-ladder results of \citet{sandage99},
\citet{parodi00} or \citet{tammann01}.

Recently, a series of papers was published by
\citet{kochanek02,kochanek02c,kochanek02b}, see also
\citet{kochanek03}, in which it is claimed that a number of
gravitational lens time-delays is compatible with the Hubble constant
from the \hst\ key project only if the mass concentration is as
compact as the light distribution ($\beta<1$ in the picture of
power-law models). Our result for \BOii\ does not confirm this view
and does not give rise to a `new dark matter problem' since it is
significantly higher than other results from lenses. Our isothermal
models lead to a value absolutely compatible with the one preferred in
\citet{kochanek02}. One could now in a similar way compare
with the lower $H_0$ results of \citet{sandage99}, \citet{parodi00} or
\citet{tammann01}. However, as long as the local determinations do not agree
with each other, we consider such an exercise to be of only limited value.

Rather than using locally measured values for $H_0$ to constrain lens
mass distribution, we prefer a more direct approach,
either using the lens effect itself (see below), or by
including additional information. The latter approach is followed by
thy `Lenses Structure and Dynamics' survey LSD
\citep{koopmans02,treu02a,treu02b,koopmans03a,koopmans03b,koopmans03c}.
The general idea is simple: The total mass within the Einstein radius
is well constrained by lensing, in contrast to the radial mass
\emph{profile}. For the given mass, the stellar velocity dispersion of the
lensing galaxy depends strongly on this profile. The more the mass is
concentrated in the centre, the deeper the potential well and the
higher the velocity dispersion has to be. With some additional
assumptions, it is indeed possible to obtain valuable constraints
which can then again be included in the lens models to determine
$H_0$.  For the important system PG~1115+080, \citet{treu02a} obtain a
power-law index of $\beta=0.65$ which increases the otherwise very low
estimate of $H_0$ for this system to a value of $59\pm10$
($1\,\sigma$). This steep mass profile is quite unusual and moves the
galaxy significantly off the fundamental plane. Nevertheless it is not
as steep as the \emph{light} profile and the result for $H_0$
correspondingly lies between the isothermal and constant $M/L$ model
results of \citet{kochanek03}. For B1608+656, \citet{koopmans03b} are
able to constrain the mass profile of the main lensing galaxy quite
well and find that it is consistent with isothermal. The resulting
Hubble constant is $75\pm7$ ($1\,\sigma)$, inconsistent with the very
low mean value of $48\pm3$ obtained by \citet{kochanek02c,kochanek02b}
for isothermal models for a number of lenses, but consistent with our
result for \BOii. However, B1608+656 has the major disadvantage of a
second lensing galaxy very close to the primary. The two galaxies
might even be interacting which could cause complicated deviations
from usual galaxy mass distributions.

These examples show that the general picture of lensing constraints for
$H_0$ and mass profiles is not as consistent as it appeared not long ago
\citep[e.g.][]{koopmans99}. 
In our opinion the only way to resolve the discrepancies is to try and use the
best available constraints for all applicable lens systems individually. Our
work on \BOii\ shows this for one example, although there are still many open
questions.

Our alternative way of constraining mass profiles relies on the lens effect
alone and has the advantage that additional astrophysics (galactic dynamics in
LSD work) and corresponding additional model assumptions are
not needed, keeping the method simple and clean.
It is easy to understand that lenses showing only two or four images of one
compact source cannot provide sufficient information to constrain the mass
profile tightly, especially since quads usually have all images close to the
Einstein ring so that they probe the potential at only one radius.
Far better suited are systems with extended sources or at least with
substructure in the compact images. \BOii\ is a good example which offers
both. 

Using our galaxy position as an estimate, non-isothermal power-law models
could be 
constrained with the VLBI substructure. These are not the final results since
no self-consistent fitting of all available data has been performed so far,
but the weak dependency of the radial mass exponent $\beta$ on the galaxy
positions shows that the estimate is nevertheless quite good. We learned that
the mass distribution in the lens is close to isothermal (which would be
$\beta=1$) but is slightly shallower ($\beta\approx 1.04$). Preliminary
calculations with \lc\ showed that the shift in the best lens position caused
by this small deviation partly compensates for the expected scaling with
$2-\beta$ for a fixed position. We therefore do not expect very significant
changes of the result for $H_0$ from the slight non-isothermality in this
system. A very conservative error estimate disregarding this compensation
effect would be a possible 4~per cent error for $H_0$.

It has to be kept in mind, however, that all results presented so far depend
on the 
assumption of a power-law for the mass profile (and potential and deflection
angle). For more general mass models the constraints on $\beta$ are still
interesting, but they only measure the \emph{local} slope\footnote{To be more
  specific the VLBI substructure provides constraints for the
  difference of the slopes of the deflection angle at the two image
  positions.}
of the profile rather than the global mass profile.
It is expected that power-laws can be used as a very good approximation for
more general 
models for limited ranges of radii, and our model fits confirm this view in
the case of \BOii. The \lc\ results for the lens position 
presented here are therefore also valid for other mass profiles, but the value
for the Hubble constant may well have to be modified slightly.

The mass models discussed here do not include external shear because it seems
to be very small. The expected true external shear of 1--2~per
cent \citep{lehar00} would change the result of $H_0$ by about the same amount
at most. More difficult to estimate is the possible influence of differences
in the ellipticity of the inner and outer parts of the lensing galaxy which
can effectively also act as external shear.
The possible impact of the remaining higher-order aspects of the mass profile
degeneracies and the effects of multi-component mass models (bulge+disc+halo)
are currently under investigation, especially the aspect of what apparent
isothermality in the central part of the galaxy, which should be dominated by
the luminous matter in the bulge, means for the true mass distribution.

In the future we will avoid the model fitting of VLBI components which are
then used as constraints and instead use \lc\ itself also on the existing VLBI
data that show a wealth of structure in the images (see
Fig.~\ref{fig:comp}). To be able to use non-isothermal models with \lc, we
developed the new method \lentil\ for the inversion of the lens equation. With
these two methods combined, a simultaneous fit of medium resolution VLA data
(sensitive to the lens position) and the existing high resolution VLBI data of
\citet{biggs02}, which are more sensitive to the radial mass profile, will not
only improve the results for all parameters 
but will, by using self-consistent lens models, also provide realistic error
estimates including all uncertainties.

To be able to use the 8.4\,GHz data set it will be necessary to include the
extinction and scatter broadening in the A image in the models. This is
possible because the 
15\,GHz data are only little affected by extinction so that the combination of
the two 
allows estimates of this effect.
Since the VLBI constraints for the lens models are mainly given by the
positions of subcomponents but not by their size, which is relatively
independent of the lens model, the different sizes of components in A and B
can be used to obtain better constraints on the scattering measure than the
simple estimates from \citet{biggs02}.

The application of \lc\ to the VLBI data will also show whether any
significant amount of substructure in the mass distribution of the lens is
needed to explain all features of the jet. The first results from
\citet{biggs02} seem to be compatible with no substructure, but a quantitative
analysis can be done in the future. If significant substructure effects are
present, it will also be necessary to correct the medium resolution data for
them before using \lc\ to determine the galaxy position. Clumps of matter
close to one of the images could change the magnification significantly which
would then mimic a different lens model. The analysis of the VLBI
data will show whether this may be the explanation for the (compared with
some other lenses) relatively high result for $H_0$.
\lc\ can also be used to study
substructure from VLBI observations of other lenses. It will then be possible
to do an analysis similar to e.g.\ \citet{metcalf02} but
without any assumption about the true source structure.

To improve the medium resolution side, we recently performed observations of
\BOii\ at 15\,GHz using the fibre link 
between the VLA and the VLBA telescope Pie~Town. This combination provides a
resolution 
1.5--2 times better than the VLA alone. The long track observation of the total
accessible hour angle range of 14~hours also improves the sensitivity
which in combination with the improved resolution can reduce both statistical
and (even more important) systematical errors. Simulations performed before
the observations showed that an accuracy of about 1--2$\,\mas$ for the lens
position should be achievable. This would in comparison to the existing
data be an improvement of a factor of $\sim 5$. The remaining uncertainty in
$H_0$ from the lens position alone will then shrink to below 2~per~cent. 
Possible uncertainties caused by calibration errors will also be reduced
significantly. 
This new data set is currently being calibrated and will be analysed with \lc\
soon, alone and in combination with the VLBI data.

As a secondary result of \lc\ we presented maps of the brightness distribution
of the source as well as improved lens plane maps, both produced with a method
newly developed together with \lc\ in \papi.
This allowed the first
view of \BOii's source as it would be seen without the distortion of the lens,
but with improved resolution.
The ring itself is caused by a bending jet which crosses the tangential
caustic of the lens. In the future we will improve these results and also
produce source plane polarization maps. The ring shows an interesting radial
polarization pattern \citep{biggs02} and the data will, when projected to the
source plane, give us a very detailed and magnified view of the polarization
structure in the jet of the source.

Finally we used \lc\ to investigate the puzzling changes of the flux density
ratio A$/$B with frequency. To test the two theories of either scattering
induced extinction caused by the ISM in the lensing galaxy or of an effective
shift of the source with frequency, we compared our best \lc\ models, which
were fitted to 15\,GHz data where the fluxes are expected to be unaffected by
propagation 
effects, with MERLIN data at 5\,GHz taking into account a possible relative
shift and extinction in the 5\,GHz data. Although the 5\,GHz data set has some
calibration problems, the evidence for a significant extinction ($\sim
25$~per~cent) of the A component at 5\,GHz is very strong. The relative shift
of the source between these two frequencies seems to be negligible.
This work will be extended in the future by comparing
medium resolution data sets at different frequencies in order to study the
propagation effects in the lensing galaxy 
in more detail. Observations with MERLIN at different frequencies will allow
us to measure the position dependent Faraday rotation and depolarisation,
providing invaluable information about the ISM of the lensing galaxy.

A direct measurement of the galaxy position in \BOii\ by \hst/ACS
observations 
will be available soon. The formal uncertainty of this measurement may be
larger than the \lc\ model constraints but it will be an absolutely
independent and therefore complementary direct measurement.
The comparison will either confirm the lens models or give information about
deviations of the true mass distribution from the relatively simple models.
We will then reach the point were the uncertainty in $H_0$ is no longer
dominated by the lens models but by the time-delay uncertainties themselves. A
reanalysis of existing monitoring data \citep{biggs99,cohen00} and new
monitoring campaigns will therefore allow further improvements.
With all the constraints available now or soon and the simplicity of the
possible mass models, \BOii\ has the potential to lead to the most robust
measurement of $H_0$ of all time-delay lenses.

\section*{Acknowledgments}

The authors would like to thank the anonymous referee for a very helpful
report and the Royal Astronomical Society for covering the cost of
the colour figure.

O.W. was funded by the `Deutsche Forschungsgemeinschaft', reference
no.\ Re~439/26--1 and 439/26--4; European Commission, Marie Curie Training
Site programme, under contract no.\ HPMT-CT-2000-00069 and TMR Programme,
Research Network Contract ERBFMRXCT96-0034 `CERES'; and by the
BMBF/DLR Verbundforschung under grant 50~OR~0208.

\newcommand{\mnras}{\mbox{MNRAS}}
\newcommand{\apj}{\mbox{ApJ}}
\newcommand{\apjs}{\mbox{ApJS}}
\newcommand{\aj}{\mbox{AJ}}
\newcommand{\aaps}{\mbox{A\&AS}}
\newcommand{\aap}{\mbox{A\&A}}

\appendix

\section{Using the subcomponents as model constraints\label{sec:app matrix}}

Applying the lens equation \eqref{eq:lens} for two nearby positions
gives to first order:
\begin{equation}
\Delta\vc\zs = \mat{M}^{-1} \Delta\vc z
\label{eq:lens diff}
\end{equation}
Here $\mat{M}$ denotes the magnification matrix
\begin{equation}
(\mat{M}^{-1})_{ij} = \delta_{ij} - \frac{\upartial
  \alpha_i}{\upartial z_j} \rtext{.}
\nt
\end{equation}
Clearly any information about the small scale structure of the images
A and B (flux density ratio, subcomponent positions and shapes) can
only provide constraints for the \emph{relative} magnification matrix
%$\ma^{-1}\mb$.
$\ma^{-1}\,\mb^{\vphantom{\mymathstrut}}$.

\subsection{Relative Positions}
\label{sec:app vec}

It would be possible to use the subcomponents of A and B as
images of two independent sources to constrain the models in the usual
way. Since the separation of the subcomponents is much smaller than the
typical scale of the system, however, these data can only constrain the first
derivative of the deflection angle, i.e.\ the magnification matrix. A
different approach to constrain this matrix directly is therefore more
appropriate. 

Given a model value for the relative position vector of $\rs$ in the
source plane, the 
corresponding vectors in the lens plane are given by Eq.~\eqref{eq:lens
  diff}, leading to a contribution to $\chi^2$ of
\begin{equation}
\chi^2 = \sum_{i=\rmn{A,B}} \transp{\bigl(\vc r_i-\mat{M}_i\rs\bigr)}
\,\mat{C}_i^{-1} \, \bigl(\vc r_i-\mat{M}_i\rs\bigr)
\nt
\end{equation}
with the measured lens plane vectors of $\ra$, $\rb$ and corresponding
covariance matrices $\mat{C}\sub A$ and $\mat{C}\sub B$.
This equation defines an ordinary linear least squares problem leading to a
weighted mean of the backprojected $\vc r_i$ for the solution:
\begin{equation}
\rs = \left(\sum_{i=\text{A,B}} \mat{M}_i \mat{C}_i^{-1}
  \mat{M}_i\right)^{-1} \, \sum_{i=\text{A,B}} \mat{M}_i
  \mat{C}_i^{-1}\, \vc r_i
\nt
\end{equation}

\subsection{Shapes and sizes}
\label{sec:app shapes}

With the common polar ellipticity parameters, fits can only be done
numerically and are potentially very unstable. The following Cartesian
formalism allows a direct 
analytical fit for the elliptical shape of the source to minimize the
deviations from the observed images.
An ellipse with major and minor axis $a$ and $b$ with direction of the
major axis $\phi$ (mathematical sense) can be described by the matrix
$\mat{E}$. For all points $\vc x$ located on the ellipse, the following
equation holds:
\begin{gather}
\transp{\vc x} \mat{E}^{-1} \vc x = 1 \nt \\[1ex]
\mat{E} = \begin{pmatrix} R_+ + R_- \cos2\phi & R_- \sin2\phi \\
                          R_-\sin2\phi & R_+ - R_-\cos2\phi
                        \end{pmatrix} \nt
\end{gather}
\begin{align}
R_+ &= \frac{a^2+b^2}{2} \nt \\
R_- &= \frac{a^2-b^2}{2} \nt
\end{align}
The astronomical position angle (north through east) is
related by $\mbox{p.a.}=\upi/2-\phi$.
Transformation between source and lens plane is linear for the
ellipticity parameters $E_{ij}$:
\begin{align}
\mat{E}_{\rmn{s}} &= \ma^{-1}\mat{E}_{\rmn{A}}\ma^{-1} \nt \\
&= \mb^{-1}\mat{E}_{\rmn{B}}\mb^{-1} \nt
\end{align}
We can therefore use the same simple general linear least squares
formalism as in Appendix~\ref{sec:app vec} with transformation
matrices for the three-dimensional ellipticity vector
$(R_+,R_-\cos2\phi,R_-\sin2\phi)$.
In this way the best source plane ellipticity matrix/vector and the residual
contributions can be calculated analytically. The Cartesian approach also
avoids singularities and degeneracies for certain shapes, e.g.\ for circular
sources.

\bsp

\end{document}